\documentclass[]{aa}  
\usepackage{amsmath}
\usepackage[colorlinks,breaklinks]{hyperref}
\hypersetup{linkcolor=blue,citecolor=blue,filecolor=black,urlcolor=blue}
\usepackage{xcolor}
\usepackage{amssymb}

\newcommand{\kms}{km~s$^{-1}$}

\newcommand{\SiII} {\ion{Si}{ii}}
\newcommand{\SiIIs} {\SiII\ $\lambda6355$}

\usepackage{xspace}
\usepackage{graphicx}
\usepackage{txfonts}
\usepackage{pdflscape}
\usepackage{comment}
\usepackage{orcidlink}
\usepackage[switch]{lineno}
\usepackage{hyperref}

\newcommand{\customcitecolor}[2]{%
    \begingroup
    \hypersetup{citecolor=#1}%
    \cite{#2}%
    \endgroup
}

\newcommand{\customcitecolornew}[2]{%
    \begingroup
    \hypersetup{citecolor=#1}%
    \citealt{#2}%
    \endgroup
}

\begin{document}

%
%
    \title{ZTF SN Ia DR2 follow-up: Exploring the origin of the Type Ia supernova host galaxy step through Si II velocities}
    \author{U. Burgaz \inst{1}\orcidlink{0000-0003-0126-3999}
    \and K. Maguire\inst{1}\orcidlink{0000-0002-9770-3508}
    \and L.~Galbany\inst{2,3}\orcidlink{0000-0002-1296-6887}
    \and M.~Rigault\inst{4}\orcidlink{0000-0002-8121-2560}
    \and Y.-L.~Kim\inst{5}\orcidlink{0000-0002-1031-0796}
    \and J. Sollerman\inst{6}\orcidlink{0000-0003-1546-6615}
    \and T.~E.~Muller-Bravo\inst{1,7}\orcidlink{0000-0003-3939-7167}
    \and M.~Ginolin\inst{4}\orcidlink{0009-0004-5311-9301}
    \and M.~Smith\inst{8}\orcidlink{0000-0002-3321-1432}
    \and G. Dimitriadis\inst{8}\orcidlink{0000-0001-9494-179X}
    \and J.~Johansson\inst{9}\orcidlink{0000-0001-5975-290X}
    \and A.~Goobar\inst{9}\orcidlink{0000-0002-4163-4996}
    \and J. Nordin\inst{10}\orcidlink{0000-0001-8342-6274}
    \and P.~E.~Nugent\inst{11,12}\orcidlink{000-0002-3389-0586}
    \and J.~H.~Terwel\inst{1}\orcidlink{0000-0001-9834-3439}
    \and A. Townsend\inst{10}\orcidlink{0000-0001-6343-3362}
    \and R.~Dekany\inst{13}\orcidlink{0000-0002-5884-7867}
    \and M.~J.~Graham\inst{14}
    \and S.~L.~Groom\inst{15}\orcidlink{0000-0002-9998-6732}
    \and N.~Rehemtulla\inst{16,17,18}\orcidlink{0000-0002-5683-2389}
    \and A.~Wold\inst{15}\orcidlink{0000-0001-5668-3507}
    }

   \institute{School of Physics, Trinity College Dublin, College Green, Dublin 2, Ireland\\
            \email{burgazu@tcd.ie}
            \and Institute of Space Sciences (ICE-CSIC), Campus UAB, Carrer de Can Magrans, s/n, E-08193 Barcelona, Spain\   
            \and Institut d'Estudis Espacials de Catalunya (IEEC), 08860 Castelldefels (Barcelona), Spain\ 
            \and Université Lyon, CNRS, IP2I Lyon/IN2P3, UMR 5822, F-69622, Villeurbanne, France\
            \and Department of Astronomy \& Center for Galaxy Evolution Research, Yonsei University, Seoul 03722, Republic of Korea\
            \and Oskar Klein Centre, Department of Astronomy, Stockholm University, SE-10691 Stockholm, Sweden\
            \and Instituto de Ciencias Exactas y Naturales (ICEN), Universidad Arturo Prat, Chile\
            \and Department of Physics, Lancaster University, Lancaster LA1 4YB, UK\
            \and The Oskar Klein Centre, Department of Physics, Stockholm University, SE-10691 Stockholm, Sweden\
            \and Institut für Physik, Humboldt-Universität zu Berlin, Newtonstr. 15, 12489 Berlin, Germany\
            \and Lawrence Berkeley National Laboratory, 1 Cyclotron Road MS 50B-4206, Berkeley, CA, 94720, USA\
            \and Department of Astronomy, University of California, Berkeley, 501 Campbell Hall, Berkeley, CA 94720, USA\  
            \and Caltech Optical Observatories, California Institute of Technology, Pasadena, CA  91125\
            \and Division of Physics, Mathematics and Astronomy, California Institute of Technology, Pasadena, CA, 91125, USA\ 
            \and IPAC, California Institute of Technology, 1200 E. California Blvd, Pasadena, CA 91125, USA\ 
            \and Department of Physics and Astronomy, Northwestern University, 2145 Sheridan Road, Evanston, IL 60208, USA\
            \and Center for Interdisciplinary Exploration and Research in Astrophysics (CIERA), 1800 Sherman Ave., Evanston, IL 60201, USA\
            \and NSF-Simons AI Institute for the Sky (SkAI), 172 E. Chestnut St., Chicago, IL 60611, USA\
            }
   
\titlerunning{Exploring the origin of the Type Ia supernova host galaxy step through Si II velocities}
\authorrunning{U. Burgaz, et al.}

\date{}

  \abstract
  {The relationship between Type Ia supernovae (SNe Ia) and their host galaxy stellar masses has been well documented. In particular, Hubble residuals display a distinct luminosity shift based on host mass, known as the mass step. This effect is widely used as an extra correction factor in the standardisation of SN Ia luminosities. In this study, we investigate the Hubble residuals and the mass-step of normal SNe Ia, in the context of \SiIIs\ velocities, using 277 normal SNe Ia that are near-peak from the second data release (DR2) of the Zwicky Transient Facility (ZTF). We divide the sample into high-velocity (HV) and normal-velocity (NV) SNe Ia, separated at 12,000 \kms, resulting in 70 HV and 207 NV objects. We then explore potential environmental and/or progenitor-related effects by investigating the \SiIIs\ velocities with parameters such as the light-curve stretch $x_{1}$, colour $c$, and host galaxy properties. Although we only find a marginal difference between the Hubble residuals of HV and NV SNe Ia, the NV mass step is $0.149 \pm 0.024$ mag ($6.3\sigma$). The HV mass step is smaller, $0.046 \pm 0.041$ mag ($1.1\sigma$), and consistent with zero. The difference between the NV and HV mass steps is modest, at $\sim2.2\sigma$. Moreover, the clearest subtype difference appears for SNe in central regions ($d_{DLR}<1$), where NV SNe Ia show a strong mass step, whereas HV SNe Ia are consistent with no step, yielding a $3.1$–$3.6\sigma$ difference between NV and HV SNe Ia. We observe a host-colour step for both subtypes. NV SNe Ia show a step of $0.142 \pm 0.024$ mag ($5.9\sigma$), while HV SNe Ia show a step of $0.158 \pm 0.042$ mag ($3.8\sigma$), where the HV SNe Ia step appear larger, but the significance is reduced because of the smaller sample size. Overall, the NV and HV colour steps are statistically consistent. HV SNe Ia also show modest ($\sim2.5$–$3\sigma$) steps in certain subsets, such as those in outer regions ($d_{DLR}>1$), whereas NV SNe display stronger environmental trends. Our results indicate that, NV SNe Ia appear more environmentally sensitive, particularly in central, likely metal-rich and older regions, while HV SNe Ia show weaker and subset-dependent trends. This suggests that applying a universal mass-step correction could introduce biases, and that incorporating refined classifications and/or environment-dependent factors, such as location within the host, may improve future cosmological analyses beyond the standard $x_{1}$ and $c$ cuts.}


   \maketitle

\section{Introduction}
\label{sec:HVNVintro}
Type Ia supernovae (SNe Ia) are proven to be reliable distance estimators and are used as `standardisable candles' to measure extragalactic distances \citep{phillips1993,hamuy1995,riess1996}. This property has made them crucial in cosmology, leading to the discovery of the accelerating expansion of the Universe \citep{riess1998,perlmutter1999} and the measurement of the Hubble constant $H_0$ \citep{riess2009,Freedman2019,riess2022,Galbany2023}. 

SNe Ia exhibit an observed scatter of about 0.4 mag in M$_{B}$ \citep{Branch1993}. However, this dispersion can be significantly reduced to approximately 0.15 mag by applying empirical corrections based on the `brighter-slower' and `brighter-bluer' correlations \citep{phillips1993,riess1996,tripp1998}. These corrections are based on light curve properties, specifically stretch $x_1$ and colour $c$, which are commonly determined using light curve fitting methods such as SALT2 \citep{Guy2007,Guy2010}, SALT3 \citep{Kenworthy2021,Taylor2023}, SNooPy \citep{Burns2011, Burns2014}, and \texttt{BayeSN} \citep{Mandel2022, Grayling2024}. However, despite the improved scatter, an unknown source of uncertainty still remains, even after accounting for observational errors or theoretical limitations.

SNe Ia are widely accepted as the thermonuclear explosions of carbon–oxygen (C/O) white dwarfs (WDs) in binary systems \citep{hoyle1960, Nugent2011, maoz2014}. However, their progenitors and underlying explosion mechanisms remain not fully understood \citep[e.g.][]{maeda2016,jha2019,Ruiter2025}. Currently, there are two prevailing progenitor models for SNe Ia: the single-degenerate (SD) model \citep{whelan1973, nomoto1997}, in which a carbon–oxygen (C/O) white dwarf (WD) accretes material from a non-degenerate companion, such as a main-sequence star or an evolved red giant, and the double-degenerate (DD) model \citep{iben1984, webbink1984}, which involves the merger of two WDs. There are also multiple explosion mechanisms leading to a SN Ia in the DD model, such as the double-detonation scenario \citep{fink2007, sim2010, shen2018}  and the violent merger scenario \citep{raskin2009, pakmor2013, sato2015, roy2022}.

SNe Ia show a large range of photometric and spectroscopic diversity. This variation can affect their use in precision cosmology and makes it important to investigate different subtypes. In an effort to address this, \citet{wang2009_hvnv} introduced a classification system for `normal' cosmologically useful SNe Ia based solely on the velocity of the \SiIIs\ absorption feature around peak brightness. In this scheme, SNe Ia with expansion velocities ($v_{\mathrm{Si}}$), above approximately 12,000 km s$^{-1}$ are categorized as high velocity (HV) SNe Ia, while those with lower velocities fall into the normal velocity (NV) group. This variation in velocity is also suggested to be independent of the evolution of the light curve and it was suggested that the variation could result from differences in viewing angle \citep{maeda2010}. Studies also suggest that HV SNe Ia exhibit distinct characteristics, including differences in light curve evolution \citep{wang2008, wang2009_hvnv, burgaz2021}, a shorter time interval from explosion to peak brightness \citep{ganeshalingam2011}, and intrinsically redder \textit{B-V} colours \citep{pignata2008}. Additionally, HV SNe Ia may follow different extinction laws \citep{wang2009_hvnv, foleykasen2011}. Contradicting previous studies where HV SNe Ia were proposed to be found mainly in massive galaxies \citep{wang2013,pan2015,pan2020,Dettman2021}, \citet{burgaz2025b} demonstrated in a volume-limited sample that HV SNe Ia do not necessarily favour high-mass galaxies but are indeed also found in low-mass galaxies at a similar rate as NV SNe Ia. They further suggested that rather than representing a distinct population, HV SNe Ia may instead be part of a continuous distribution, with variations potentially arising from different explosion mechanisms or variations within a type of explosion mechanism.

A correlation between the mass of the host galaxies and the SN Ia light curve residuals after standardisation, `Hubble residuals' (HR) is widely acknowledged, where SNe Ia in high-mass galaxies on average appear brighter than low-mass galaxies after standardisation,  \citep{Kelly2010,Sullivan2010,Childress2013,ginolin2025b}. Therefore, a so-called `mass step' correction is routinely incorporated into cosmological analyses \citep{Lampeitl2010,Betoule2014,Scolnic2018,Brout2022,Rubin2023,DES2024,popovic2024}. Similar correlations are also seen between HRs and the local star formation rate \citep{Rigault2013, Kim2018}, the global specific star formation rate \citep{Sullivan2006}, the metallicity \citep{DAndrea2011,Childress2013,pan2014,Moreno2018,Irigoyen2022}, age \citep{Neill2009,Rigault2020,chung2023,Chung2025}, and the birth environments of SN Ia progenitor stars \citep{Kim2024}. However, their origin remain unclear \citep[e.g.][]{Hayden2013,Rose2019,Popovic2021,popovic2023}.

A relationship between \SiIIs\ velocities and HRs has also been suggested, with \citet{Siebert2020} highlighting a possible trend at the 2.7$\sigma$ level. They observed a 0.091 $\pm$ 0.035 mag difference in HR when splitting the sample into HV and LV at the median velocity (11,000 \kms), indicating a potential trend. They also noted a \SiIIs\ velocity difference of 980 $\pm$ 220 \kms\ between the positive and negative HRs. However, a similar study from \citet{Dettman2021}, using the Foundation SN sample reported a not significant 160 $\pm$ 230 \kms\ difference between their positive and negative HRs, along with 0.015 $\pm$ 0.049 mag difference between the weighted averages of the HRs of their HV and NV samples (separated at 11,800 \kms). A recent study from \citet{Pan2024}, found a 0.044 $\pm$ 0.023 mag difference at 1.9$\sigma$ significance between the weighted averages of the HRs of their HV and NV samples (separated at 12,000 \kms). Therefore, there remains uncertainty of the presence and strength of these relationships.

In this study, our objective is to investigate the Hubble residuals and the host galaxy mass step of NV and HV SNe Ia using the SN Ia data from the second data release \citep[DR2;][]{rigault2025a,smith2025} of the Zwicky Transient Facility \citep[ZTF;][]{Bellm2019,Graham2019,Masci2019,Dekany2020}, collected during the first three years of the survey from 2018 to 2020. In Section~\ref{sec:datamethod}, we present the data and the sample selection. Section~\ref{sec:results} discusses the Hubble residuals and the mass step from the point of HV and NV SNe Ia by comparing both sample and applying parameter analysis. The discussions and conclusions are presented in Section~\ref{sec:discussion} and Section~\ref{sec:conclusions}, respectively.

\section{Data}
\label{sec:datamethod}
In this section we present the origin and selection criteria of our SN Ia and host galaxy sample. In Section~\ref{sec:SNIadata} we detail the SN Ia data used in this work and the applied selection criteria, while Section~\ref{sec:redshifts} outlines the redshift sources used in this study. Section~\ref{sec:hostdata} presents the details for the host galaxy data.

\subsection{SN Ia data}
\label{sec:SNIadata}
The ZTF SN Ia DR2 sample offers a comprehensive dataset consisting of 3628 spectroscopically confirmed SNe Ia. According to survey simulations, the ZTF DR2 sample is considered complete for non-peculiar SNe Ia up to a redshift of $z \leq$ 0.06 \citep{Amenouche2025}, which includes 1584 SNe Ia. This completeness is further supported by the colour distribution \citep{ginolin2025a}, which aligns well with the intrinsic properties of the SN Ia population. 

\citet{burgaz2025a} provide a detailed analysis of the spectral diversity of the volume-limited ZTF SN Ia DR2 sample at maximum light ($-$5 d $\leq t_0 \leq$ 5 d). The sample includes 482 SNe Ia, with most optical spectra obtained using the low-resolution integral field spectrograph, the SED Machine \cite[SEDm,][]{blagorodnova2018, rigault2019, kim2022}, on the 60-inch telescope at Mount Palomar Observatory. For this study of the connection between the spectral velocities and Hubble residuals of the normal cosmologically useful SNe Ia, we restrict our sample to those in \citet{burgaz2025a} and put a further constraint that SNe Ia must also be classified as `normal' SNe Ia in \citet{burgaz2025a}. This excludes other subtypes such as 91T/99aa-like \citep{filippenko1992b,garavani2004}, 86G-like \citep{phillips1987}, 04gs-like \citep{burgaz2025a}, 91bg-like SNe Ia \citep{filippenko1992a}, 02cx-like SNe Ia \citep{li2003}, SNe Ia-CSM \citep{Benetti2006,Sharma2023}, 03fg-like SNe Ia \citep{Howell2006} and 02es-like \citep{ganeshalingam2012} SNe Ia. These peculiar objects in the ZTF sample are studied in detail and presented in \citet{dimitriadis2025}. 

\begin{table}
\centering
\caption{Summary of the cuts applied.}
\label{tab:cuts}
\begin{tabular}{l c c}
\hline\\[-0.5em]
     & Criteria & No. of SNe \\[0.15em]
    \hline\\[-0.8em]
    \hline\\[-0.5em]
Full ZTF SN Ia DR2 & & 3628\\[0.30em]
with redshift & $z$ $\leq$ 0.06 & 1584\\[0.30em]
with max.~light spectrum\tablefootmark{a} & $-$5 d $\leq$ $t_0$ $\leq$ 5 d  & 482\\[0.30em]
with assigned host & & 477 \\[0.30em]
with normal subtype SN Ia & & 295 \\
with Hubble Residual & & 277\\
    \hline\\[-0.5em]  
\end{tabular}
\tablefoot{
\tablefoottext{a}{see \citet{burgaz2025a} for details of sample selection.}\\
}
\end{table}

\begin{figure*}
\centering
\includegraphics[width=\linewidth]{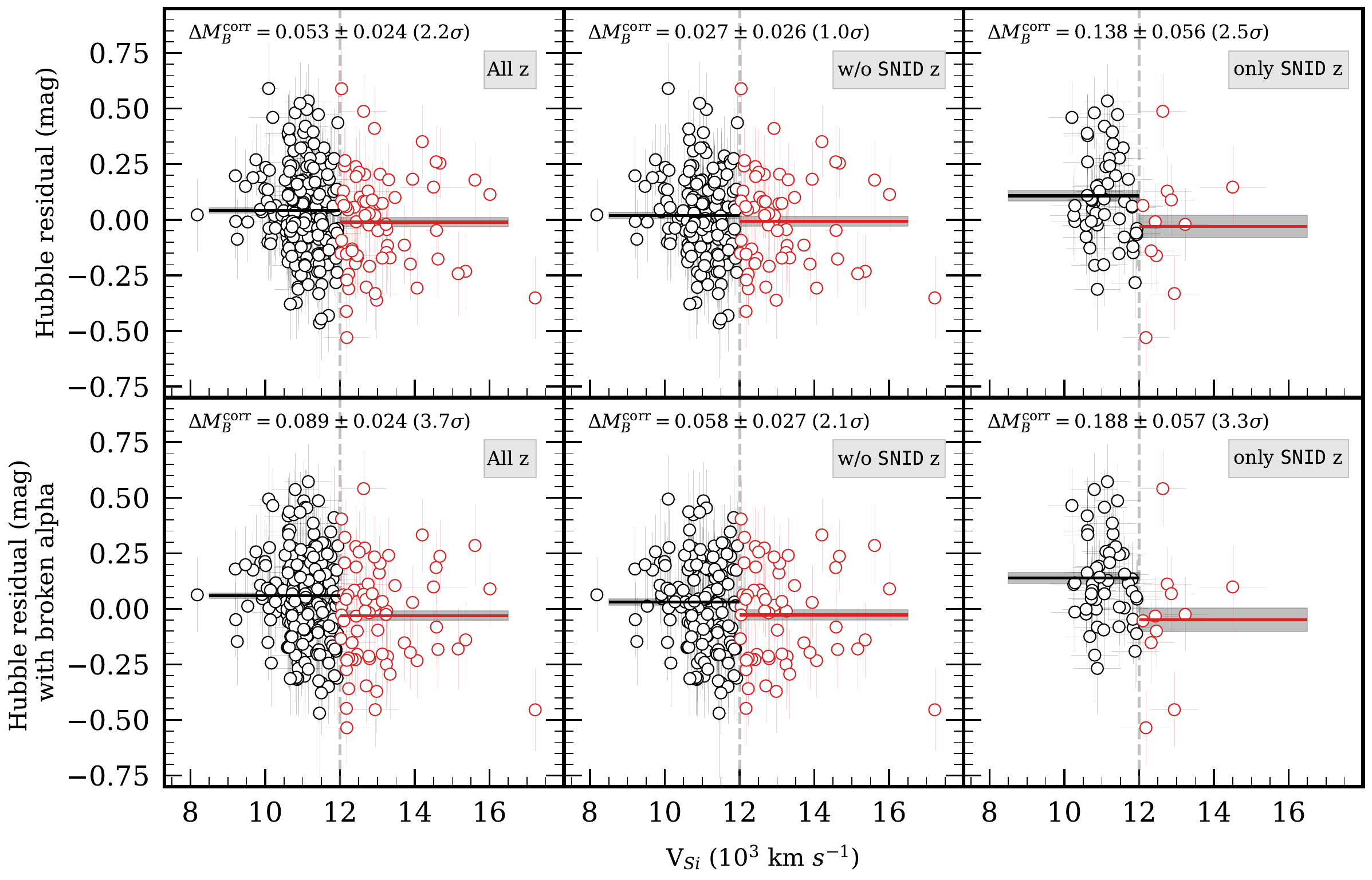}
\caption{The Hubble residuals as a function of \SiII\ $\lambda$6355 velocity ($v_{\mathrm{Si}}$). \textbf{Top Left:} Hubble residuals for NV (black points) and HV (red points) are shown with all redshift sources included (see: Section \ref{sec:SNIadata} for the redshift sources). \textbf{Top Middle:} Same as left but without the template matched redshift sources. \textbf{Top Right:} Same as left but with only the template matched redshift sources. \textbf{Bottom Left:} Hubble residuals with broken alpha from \citet{ginolin2025b} for all redshift sources included sample. \textbf{Bottom Middle:} Same as bottom left but without the template matched redshift sources. \textbf{Bottom Right:} Same as bottom left but with only the template matched redshift sources. In all panels, horizontal black and red lines represent the weighted mean of the Hubble residuals in the NV and HV bins, respectively, where the shaded regions shows the one-sigma uncertainty of the weighted means. The vertical grey dashed line represents the criterion ($v_{\mathrm{Si}}$ = 12,000 \kms) used to separate the HV and NV SN Ia samples. The difference of the weighted averages of HV and NV samples, along with the significances are shown in each panel.}
\label{fig:HR_vel_12000}
\end{figure*}

Some cosmological studies do include 91T/99aa-like SNe Ia but these events might introduce a systematic bias \citep{scalzo2012, yang2022, chakraborty2023}. In this study, since the NV and HV classifications apply only to normal SNe Ia \citep{wang2009_hvnv}, we restrict our analysis to this subset. However, to examine the potential impact of including subtypes that are close to normal SNe Ia (e.g. brighter 99aa-like and the fainter 04gs-like SNe), we explore sample selection effects in Appendix~\ref{appASample}.

Our analysis compares velocities to light-curve properties, including HRs and light curve parameters ($c$ and $x_1$). The light curves in our sample were fit as part of the ZTF SN Ia DR2 analysis, using the  \texttt{SALT2} \citep{Guy2007} light-curve fitter, retrained in \citet{Taylor2021} as `SALT2-T21' from the \texttt{sncosmo} package\footnote{\url{https://sncosmo.readthedocs.io/en/stable/about.html}} (see \customcitecolornew{red}{smith2025} for more detail). \citet{ginolin2025a,ginolin2025b} presented a detailed analysis of the SN Ia standardisation process using the volume-limited ZTF SN Ia DR2 sample. They introduced both linear and broken-alpha standardisations, with the latter allowing the stretch–luminosity coefficient alpha to take two different values for low and high stretch regimes, reflecting the observed non-linearity in the stretch-residual relation. In this work, we use HRs (corrected for $x_1$ and $c$), both with and without broken-alpha standardisation from \citet{ginolin2025b}, and apply their slightly stricter selection criteria to our sample, including a narrower cut on $x_1$ (from $\in[-4,4]$ in \citet{burgaz2025a} to $\in[-3,3]$) and an additional SALT2 light-curve fit probability requirement (greater than 10$^{-7}$, which was not present in \citet{burgaz2025a}), and a tighter constraint on the uncertainty in $c$ of $\leq$ 0.1 compared to $\leq$0.35 in \citet{burgaz2025a}. One SN remains as an outlier with a HR of $-$0.9 mag after all these cuts and this is removed manually. This results in a final subset of 277 normal SNe Ia, with the selection criteria outlined in Table~\ref{tab:cuts}.

\subsection{Redshifts}
\label{sec:redshifts}
The majority ($\sim$65\%, 180 out of 277 SNe Ia) of the redshifts for our sample come from the Dark Energy Spectroscopic Instrument (DESI) MOST Hosts programme \citep{Soumagnac2024}. 33 SNe ($\sim$12\%) have redshifts measured from the host galaxy lines in the SNe spectra. The remaining 65 SNe ($\sim$23\%) have their redshifts determined by fitting their spectra using the Python wrapper \texttt{pysnid}\footnote{\url{https://github.com/MickaelRigault/pysnid}}, which incorporates the Supernova Identification template-matching code \citep[\texttt{SNID};][]{Blondin2007} with a typical precision of $\sigma_z \approx 10^{-3}$. For further details on the redshift determination for the ZTF SN Ia DR2 sample, see \cite{rigault2025a}.

\subsection{Host galaxy data}
\label{sec:hostdata}

\customcitecolor{red}{smith2025} outlined the approach used to identify the host galaxies of the ZTF SN Ia DR2 sample and explained how their properties are determined. All images from the Legacy Imaging Survey DR9 \citep{Dey2019}, the Sloan Digital Sky Survey (SDSS) DR17 \citep{Abdurrouf2022}, and Pan-STARRS DR2 \citep[PS1;][]{Chambers2016} were acquired in a radius of 100 kpc, centered at the SN location. Kron fluxes and the directional light radius \citep[$d_{DLR}$;][]{Sullivan2006,smith2012,gupta2016}  value for each SN were then measured with the \texttt{HostPhot}\footnote{\url{https://github.com/temuller/hostphot}} software package \citep{mullerbravo2022} by using the PS1 images, where a threshold of $d_{DLR}<7$ is used to identify a potential host galaxy. In the cases where there was more than one candidate within the chosen threshold, the galaxy with the lowest value of $d_{DLR}$ is assigned as the host. \customcitecolor{red}{smith2025} also estimated stellar masses using the method of \citet{Taylor2011} and the \texttt{P\'{E}GASE.2} model \citep{fioc1997}. Although both approaches produce similar results, especially for the high-mass galaxies, this study relies solely on the mass values obtained using \texttt{P\'{E}GASE.2}.

\section{Results}
\label{sec:results}

Our main aim is to analyze the Hubble residuals of HV and NV SNe Ia in the ZTF DR2 SN Ia sample. In Section~\ref{sec:hubbleSilVel}, we present our analysis of this sample, as well as examine how this connects to $x_1$, $c$, $d_{DLR}$, host stellar mass, and global/local colour. In Section~\ref{sec:hubbleHostMass}, we revisit the mass and colour step in SNe Ia from the HV and NV perspective. We present the effect of $d_{DLR}$ on mass and colour step in NV and HV SNe Ia in Section~\ref{sec:dDLR}.

\subsection{Hubble residuals as a function of Si II velocities}
\label{sec:hubbleSilVel}

Previous studies looking at \SiIIs\ velocities and SN Ia HRs have obtained conflicting results on the significance of their relationship \citep{Siebert2020,Dettman2021,Pan2024}. In this study, we use the HV and NV SNe Ia (separated at a velocity of 12,000 \kms) from \citet{burgaz2025a} consisting of 207 NV and 70 HV SNe Ia. In Fig.~\ref{fig:HR_vel_12000}, we present the Hubble residuals with and without the broken-alpha standardisation \citep{ginolin2025b} as a function of \SiIIs\ velocities.

The impact of only including SNe Ia with host galaxy redshifts (middle panels) or also including SN template-matched redshift sources (left panels) was investigated. We also show the sample that only has the template-matched redshift sources (right panel) to clearly show how it affects the combined sample. Typically including SNe with only template-matched redshifts increases the sample size in lower mass hosts \citep{burgaz2025b}.

\begin{figure*}
\centering
\includegraphics[width=16.7cm]{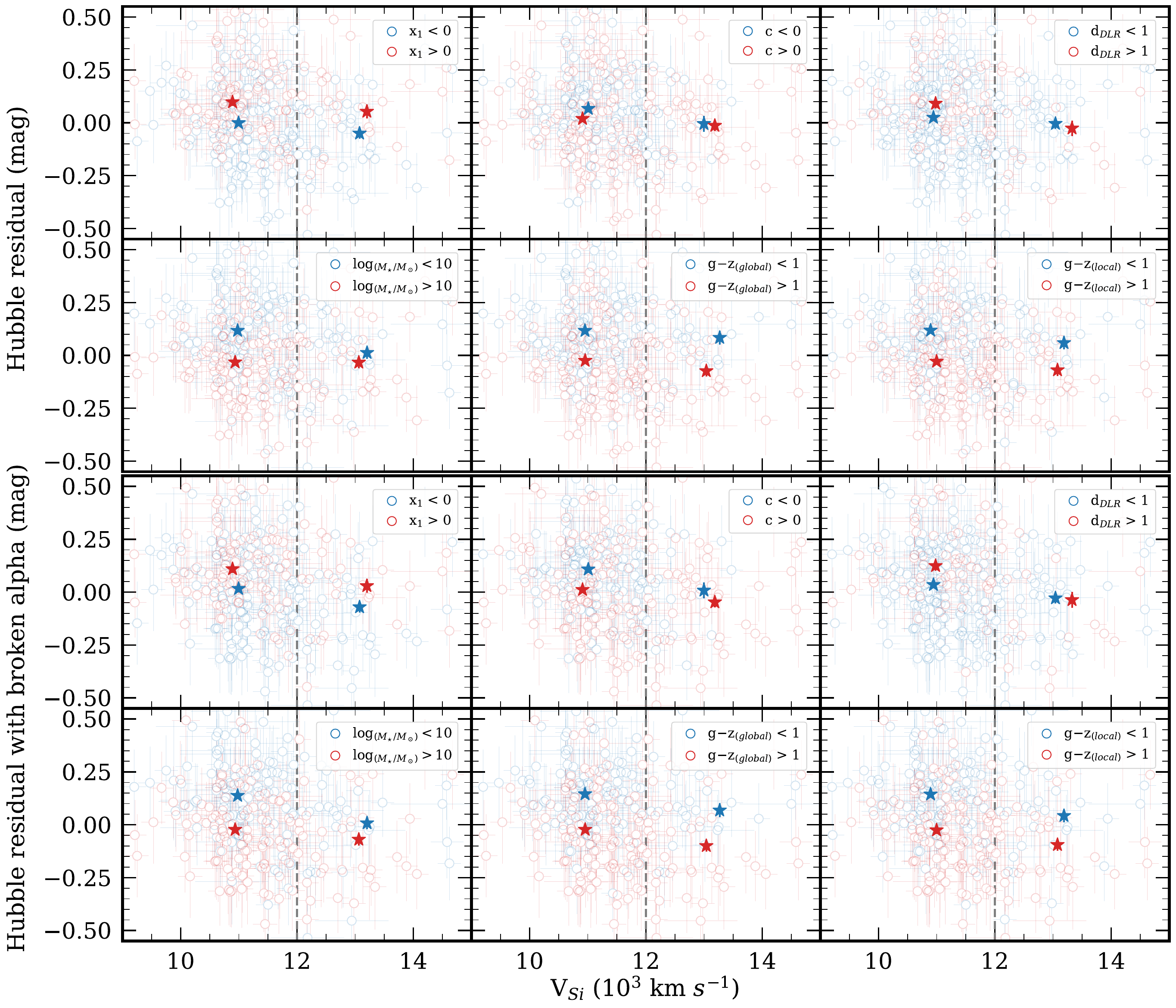}
\caption{Hubble residuals without (top two rows) and with (bottom two rows)  the broken alpha standardisation of HV and NV SNe Ia are shown as a function of \SiIIs\ velocity (all redshift sources included), separated by low and high values of $x_1$, $c$, $d_{DLR}$, host galaxy mass, global colour, and local colour. In each plot, the weighted averages (corresponding to high and low values of the parameter) for HV and NV SNe Ia are shown as stars. The vertical dashed line represents the criterion ($v_{\mathrm{Si}}$ = 12,000 \kms) used to separate the HV and NV SN Ia samples. To highlight the relative offsets, the panels here show a zoomed-in view of the Hubble residuals. The full scale version is provided in Appendix~\ref{appASample}.}
\label{fig:HR_HV_NV}
\end{figure*}

\begin{table*}
\caption{Weighted averages and significances of the Hubble residuals for the HV and NV samples across different parameters.}
\centering
\scalebox{0.95}{
\tiny
\begin{tabular}{l c c c c c c c c c} 
\hline\\[-0.5em]
 & Threshold (T) & NV$_{(<T)}$ (mag) & NV$_{(>T)}$ (mag) & HV$_{(<T)}$ (mag) & HV$_{(>T)}$ (mag) & $\Delta_{\rm NV}$ (mag) & $\Delta_{\rm NV}$ ($\sigma$) & $\Delta_{\rm HV}$ (mag) & $\Delta_{\rm HV}$ ($\sigma$) \\
\hline\\[-0.8em]
\hline\\[-0.5em]
\multicolumn{10}{c}{w/o Broken Alpha standardisation}\\
\hline\\[-0.5em]
$x_1$&0&-0.001$\pm$0.016&0.098$\pm$0.018&-0.050$\pm$0.026&0.053$\pm$0.033&-0.099$\pm$0.024&\textbf{4.1}&-0.103$\pm$0.042&2.5\\
$c$&0&0.067$\pm$0.017&0.019$\pm$0.017&-0.005$\pm$0.037&-0.013$\pm$0.024&0.048$\pm$0.024&2.0&0.008$\pm$0.044&0.2\\
$d_{DLR}$&1&0.025$\pm$0.014&0.091$\pm$0.023&-0.004$\pm$0.024&-0.026$\pm$0.037&-0.066$\pm$0.027&2.5&0.022$\pm$0.044&0.5\\
$\log({M}_\star/{M}_\odot)$&10&0.117$\pm$0.017&-0.032$\pm$0.017&0.013$\pm$0.029&-0.033$\pm$0.029&0.149$\pm$0.024&\textbf{6.3}&0.046$\pm$0.041&1.1\\
$g-z_{(global)}$&1&0.117$\pm$0.017&-0.025$\pm$0.016&0.084$\pm$0.032&-0.074$\pm$0.026&0.142$\pm$0.023&\textbf{5.9}&0.158$\pm$0.041&\textbf{3.8}\\
$g-z_{(local)}$&1&0.118$\pm$0.017&-0.029$\pm$0.017&0.059$\pm$0.030&-0.069$\pm$0.028&0.147$\pm$0.024&\textbf{6.2}&0.128$\pm$0.041&\textbf{3.2}\\
\hline\\[-0.8em]
\multicolumn{10}{c}{with Broken Alpha standardisation}\\
\hline\\[-0.5em]
$x_1$&0&0.016$\pm$0.017&0.110$\pm$0.018&-0.071$\pm$0.027&0.029$\pm$0.033&-0.094$\pm$0.025&\textbf{3.8}&-0.100$\pm$0.042&2.3\\
$c$&0&0.108$\pm$0.018&0.011$\pm$0.017&0.008$\pm$0.039&-0.047$\pm$0.025&0.097$\pm$0.025&\textbf{3.9}&0.055$\pm$0.047&1.2\\
$d_{DLR}$&1&0.035$\pm$0.014&0.125$\pm$0.024&-0.028$\pm$0.025&-0.037$\pm$0.038&-0.090$\pm$0.028&\textbf{3.2}&0.009$\pm$0.046&0.2\\
$\log({M}_\star/{M}_\odot)$&10&0.138$\pm$0.017&-0.023$\pm$0.017&0.008$\pm$0.030&-0.070$\pm$0.030&0.161$\pm$0.024&\textbf{6.6}&0.078$\pm$0.042&1.9\\
$g-z_{(global)}$&1&0.145$\pm$0.018&-0.023$\pm$0.017&0.068$\pm$0.033&-0.100$\pm$0.027&0.168$\pm$0.025&\textbf{6.9}&0.168$\pm$0.043&\textbf{4.0}\\
$g-z_{(local)}$&1&0.144$\pm$0.017&-0.025$\pm$0.017&0.042$\pm$0.031&-0.095$\pm$0.029&0.169$\pm$0.024&\textbf{6.9}&0.137$\pm$0.043&\textbf{3.3}\\
\hline\\[-0.5em]
\end{tabular}
}
\tablefoot{First column shows the parameters used in this analysis. Second column shows the threshold (T) chosen as the separation value for corresponding parameter. Columns 3–6 give the weighted averages of the HR for HV and NV sample at higher and lower than the threshold values. Columns 7–10 report the within-sample steps $\Delta_{\rm NV}\equiv{\rm NV}_{(<T)}-{\rm NV}_{(>T)}$ and $\Delta_{\rm HV}\equiv{\rm HV}_{(<T)}-{\rm HV}_{(>T)}$ with uncertainties and their significances. Values $>3\sigma$ are shown in bold.}
\label{tab:HRHVNVsigmas}
\end{table*}

We first compare the weighted average Hubble residuals between HV and NV SNe Ia. Using all redshift sources, we find a difference in the HR residuals of 0.053 $\pm$ 0.024 mag (2.2$\sigma$), while removing the template-matched sources yields 0.027 $\pm$ 0.026 mag (1$\sigma$; Fig.~\ref{fig:HR_vel_12000}). Adopting the broken-alpha of \citet{ginolin2025b} increases these differences to 0.089 $\pm$ 0.024 mag (3.7$\sigma$) and 0.058 $\pm$ 0.027 mag (2.1$\sigma$) for the same two samples, respectively. For completeness, we also show the subsample with only template matched redshifts. The differences are 0.138 $\pm$ 0.056 mag (2.5$\sigma$) without broken-alpha and 0.188 $\pm$ 0.057 mag (3.3$\sigma$) with broken-alpha standardization.

Following \citet{Siebert2020}, whose sample within four days of maximum light, was divided at the median Si II $\lambda6355$ velocity (11,000 \kms), we test the differences in HR splitting at the median velocity (11,300 \kms) for our sample (still defined within 5 days of maximum light). Using only host-galaxy spectroscopic redshifts as in \citet{Siebert2020}, we find a HR step between NV and HV of 0.052 $\pm$ 0.024 mag (2.2$\sigma$) without broken-alpha, and 0.076 $\pm$ 0.024 mag (3.1$\sigma$) with the broken-alpha standardisation. When we additionally include SNe with template matched redshifts in our sample, the HR step increases slightly to 0.073 $\pm$ 0.021 mag (3.6$\sigma$) and 0.095 $\pm$ 0.021 mag (4.5$\sigma$), respectively. Hence, our separation at 12,000 \kms\ for the HV–NV division, and the 11,300 \kms\ split adapted from the method of \citet{Siebert2020}, yield HR step measurements that are statistically consistent within $1\sigma$, regardless of the redshift source selection.

Similar to previous analyses, in all studied cases we see negative weighted averages for the HV SNe Ia, while the weighted averages of the NV SNe Ia are always positive. Inclusion of SNe Ia with template-matched redshifts in both cases (with or without the broken alpha standardisation), increases the significance of the velocity step. However, overall these are low significance differences, with only the velocity step using the broken-alpha being significant at 3.7$\sigma$ (bottom left of Fig.~\ref{fig:HR_vel_12000}). Lastly, quite similar to the results of \citet{Dettman2021}, we see an insignificant difference of 240 $\pm$ 150 \kms in the mean \SiIIs\ velocities between the positive and negative HR samples.

\subsubsection{Splits by SALT2 light-curve parameters}

Since our sample is considerably larger than previous samples, we can split both HV and NV SNe Ia for high-low $x_1$, $c$, $d_{DLR}$, host stellar mass, and global and local \textit{g-z} colour and investigate possible mechanisms that are driving the potential differences seen in Fig.~\ref{fig:HR_vel_12000}. 

In the top left panel of Fig.~\ref{fig:HR_HV_NV}, we firstly compare the HRs without a broken-alpha standardisation of the low $x_1$ ($x_1$ < 0, blue points) and high $x_1$ ($x_1$ > 0, red points) samples as a function of \SiII\ velocity. We find no significant difference between the HRs of the samples when splitting at 12,000 \kms\ (Table~\ref{tab:HRHVNVsigmas}). No significant difference is seen when the broken-alpha HRs are used (third row, left panel of Fig.~\ref{fig:HR_HV_NV}). We then examine the low and high $x_1$ samples of HV and NV SNe Ia separately. Within the NV SNe Ia (comparison between blue and red points at low velocity), we identify a $\sim$ 4$\sigma$ difference in weighted HR when split by $x_1$. This trend is slightly weaker when considering the HR with the broken alpha, but still above 3$\sigma$ significance. No significant difference in HR is seen within the HV sample. This trend in the NV sample is likely due to a remaining correlation between HR and $x_1$, also seen when the sample not split by velocity \citep{ginolin2025b}. It is likely not seen in the HV sample because of the smaller sample size compared to the NV sample. 

The SALT2 $c$ parameter shows no significant trend (all < 3$\sigma$) between the HV and NV samples with or without the broken alpha standardisation, nor within the HV and NV samples separately for the HR without broken alpha standardisation. However, a 3.9$\sigma$ difference is seen between the weighted average HR of the low $c$ NV sample and the high $c$ NV sample using the broken-alpha standardisation (Table~\ref{tab:HRHVNVsigmas}). 

\subsubsection{Splits by host-galaxy parameters}

Within the NV sample, we find a $3.2\sigma$ difference in HR between low- and high-$d_{DLR}$ subsets when using the broken-alpha standardisation, while the HV sample shows no significant trend with either method. This behaviour is similar to the SALT2 $c$ split, where a HR difference was present in the NV sample but not in the HV sample. The combined HV and NV sample also shows no trend when split at $d_{DLR}=1$, consistent with previous studies \citep{galbany2012,toy2025}.

For the host-mass split (Fig.~\ref{fig:HR_HV_NV}, bottom left panel), the low-mass NV and HV samples show differences in their weighted average HRs, with significances of 3.1$\sigma$ (with broken-alpha) and 3.8$\sigma$ (without broken-alpha). The high mass samples on the other hand, shows no significance for both standardisation models. Within the NV sample, we see a strong trend between low- and high-mass hosts, with mass steps of 6.3$\sigma$ (without broken-alpha) and 6.6$\sigma$ (with broken-alpha). By contrast, the HV sample shows no significant difference for either standardisation.

Similar to host galaxy mass, global colour (Fig.~\ref{fig:HR_HV_NV}, bottom middle panel) shows strong HR trends in the NV sample, with significances of 5.9$\sigma$ (without broken-alpha) and 6.9$\sigma$ (with broken-alpha). Unlike mass, the HV sample also shows a trend, at 3–4$\sigma$ between the low- and high-colour subsets. No significant differences are found between HV and NV at either low or high colour. Using local instead of global colour gives consistent results, with slightly higher significances (Table\ref{tab:HRHVNVsigmas}).

\subsubsection{Summary of parameter splits}

As summarised in Table~\ref{tab:HRHVNVsigmas}, for the HRs as a function of velocity, we find significant differences within the NV subsample without broken alpha standardisation when splitting the sample for $x_1$, as well as for the host parameters of stellar mass, and local and global colours. All of the parameters (both SN Ia light curves and host galaxy ones) investigated are seen to show significance differences in HRs within the NV subsample when the broken-alpha formulation is considered. We consider the stellar mass, and local and global colours has the most robust significant HR differences with the NV sample because they are seen both without and with broken alpha standardisation at $>$5.9 sigma.

For both the non-broken alpha and broken alpha analyses, we see lower level significances in the HRs for the HV subsample for the global and local galaxy colour (Table~\ref{tab:HRHVNVsigmas}), and no significant difference for global mass. Since a significant difference in HRs is seen in both the NV and HV samples (although lower for the HV sample) for both global and local colour, this suggests that the velocities measured at peak in SNe Ia are not driving this HRs difference, otherwise it would be seen in just one of the samples. However, the difference in HRs with the NV sample is very significant at $>$6 sigma for host stellar mass but a difference in HRs within the NV sample is not seen at all, suggesting that host stellar mass may be more closely linked to the SN velocity.  

We have chosen to use a velocity of 12,000 \kms\ to split our sample, following \citet{burgaz2025a}. However, the exact velocity used for this split is somewhat arbitrary, with other studies employing values of 11,800 \kms\ or 12,000 \kms. Different phase definitions and different sub-classification selections have also been used. In Appendix~\ref{appASample}, we show the results from our sample when the velocity separation is taken as 11,800 \kms and present an analysis with transitional subclasses near the normal subtype, such as 99aa-like and 04gs-like SNe Ia included to investigate the sample selection effect. We lastly investigate the phase selection effect by applying a stricter phase cut (within 3 d of peak) on our sample in Appendix~\ref{appASample}. Lowering the velocity separation threshold from 12,000 \kms\ to 11,800 \kms\ slightly enhances the significance levels when considering only normal SNe Ia. However, this increase in significance becomes negligible once 99aa-like and 04gs-like SNe are incorporated into the sample.

\subsection{Host mass and colour steps}
\label{sec:hubbleHostMass}

\begin{figure*}
\centering
\begin{minipage}{0.99\textwidth}
  \centering
  \includegraphics[width=0.99\linewidth]{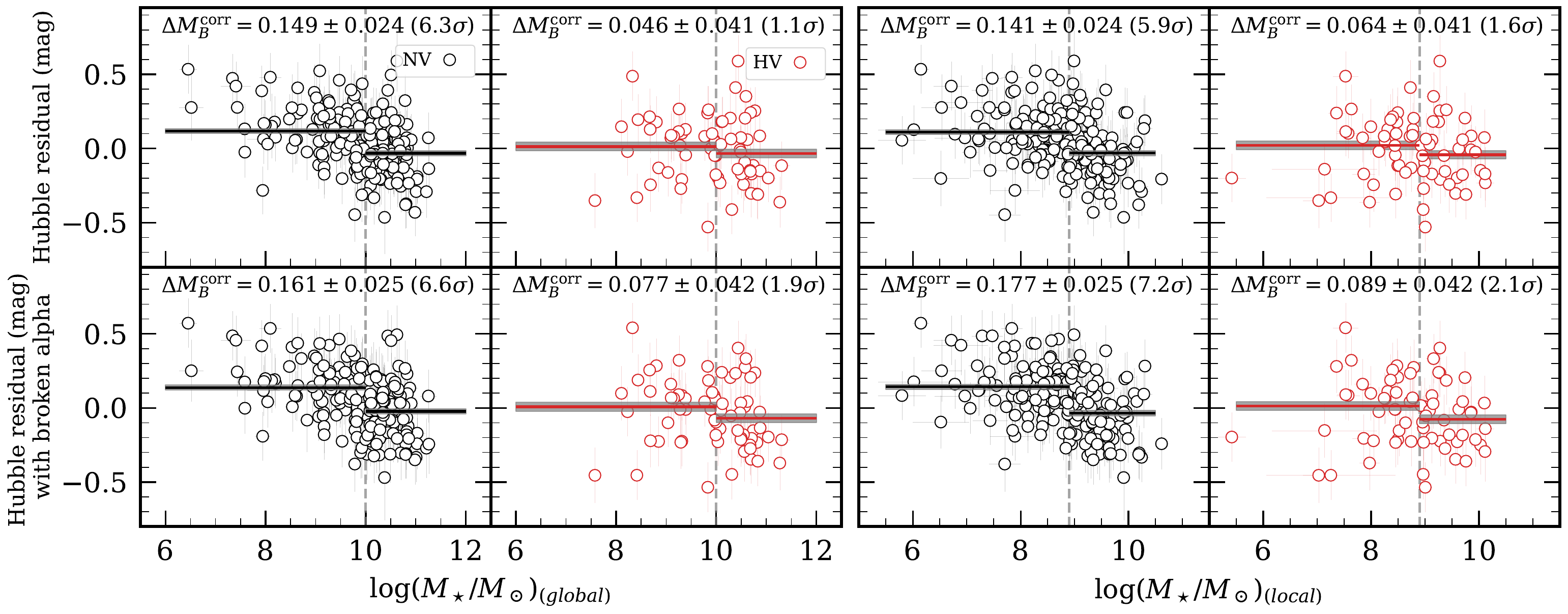}
\end{minipage}%

\begin{minipage}{0.99\textwidth}
  \centering
  \includegraphics[width=0.99\linewidth]{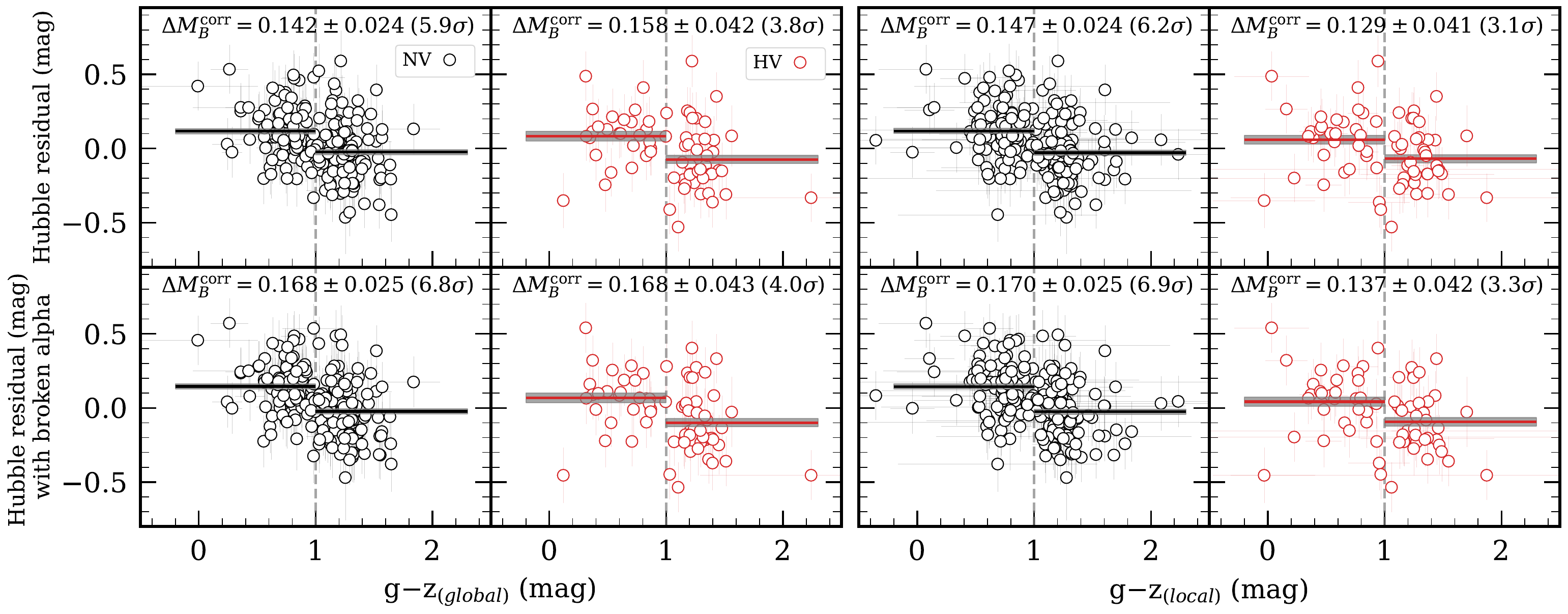}
\end{minipage}
\caption{Hubble residuals as a function of global and local host galaxy mass are shown on top two rows. The top row of plots displays the Hubble residuals without the broken-alpha standardisation, while the bottom row shows the Hubble residuals with it. In each plot, black circles represent NV SNe Ia, and red circles represent HV SNe Ia. The black and red horizontal lines show the weighted averages of the Hubble residuals for the NV and HV samples, respectively, with the shaded regions represent the one-sigma uncertainties of those averages. The vertical dashed lines at log(${M}_\star$/${M}_\odot$) = 10 show the division between low and high mass samples based on global host galaxy mass in the left panels, while the dashed lines at log(${M}_\star$/${M}_\odot$) = 8.9 show the division based on local host galaxy mass in the right panels. The difference in the weighted averages between the low and high mass samples, along with the corresponding significances, are shown in each panel. Hubble residuals as a function of global and local g$-$z colours are shown on bottom two rows. The vertical dashed line represents the criterion (\textit{g-z} = 1) used to separate the low and high colour environments.}
\label{fig:HR_mass}
\end{figure*}

\citet{ginolin2025b} demonstrated a larger HR mass-step in stretch standardisation by introducing a broken-alpha model compared to a standard alpha model. This was complemented by \citet{ginolin2025a}, focusing on SN Ia colour and dust, to explore colour standardisation. In this work, we investigate the global host-mass step separately for HV and NV SNe Ia. 

As shown in the left panels of Fig.~\ref{fig:HR_mass}, only the NV SNe Ia show a significant global mass step when the sample is split at log(${M}_\star$/${M}_\odot$) = 10, while HV SNe Ia show no mass step, both without and with a broken-alpha standardisation. For NV events we measure steps of
0.149 $\pm$ 0.024 mag (6.3$\sigma$) without broken-alpha standardisation, and 0.161 $\pm$ 0.025 mag (6.6$\sigma$) with broken-alpha standardisation. In contrast, HV SNe Ia show no significant mass step, with values of 0.046 $\pm$ 0.041 mag (1.1$\sigma$) and 0.077 $\pm$ 0.042 mag (1.9$\sigma$) under the two standardisations, respectively. The difference between the NV and HV mass steps corresponds to 0.103 mag (2.2$\sigma$) without broken-alpha, and 0.083 mag (1.7$\sigma$) with broken-alpha, respectively. Using a velocity separation of 11,800 \kms\ instead of 12,000 \kms\ yields slightly higher significances for the NV–HV mass-step difference, with values of 0.115 mag (2.6$\sigma$) without broken-alpha standardisation and 0.087 mag (1.9$\sigma$) with broken-alpha. The impact of the velocity cut choice is investigated further in Appendix~\ref{appASample}.

We also repeat the analysis using local, rather than global, host galaxy masses (right panels of Fig.~\ref{fig:HR_mass}). The NV SNe Ia again show a significant local mass step of 0.141 $\pm$ 0.024 mag (5.9$\sigma$) without broken-alpha standardisation, and 0.177 $\pm$ 0.025 mag (7.2$\sigma$) with broken-alpha. In contrast, HV SNe Ia show no significant local mass step, with values of 0.064 $\pm$ 0.041 mag (1.6$\sigma$) and 0.089 $\pm$ 0.042 mag (2.1$\sigma$), respectively.

\begin{figure*}
\centering
\includegraphics[width=\linewidth]{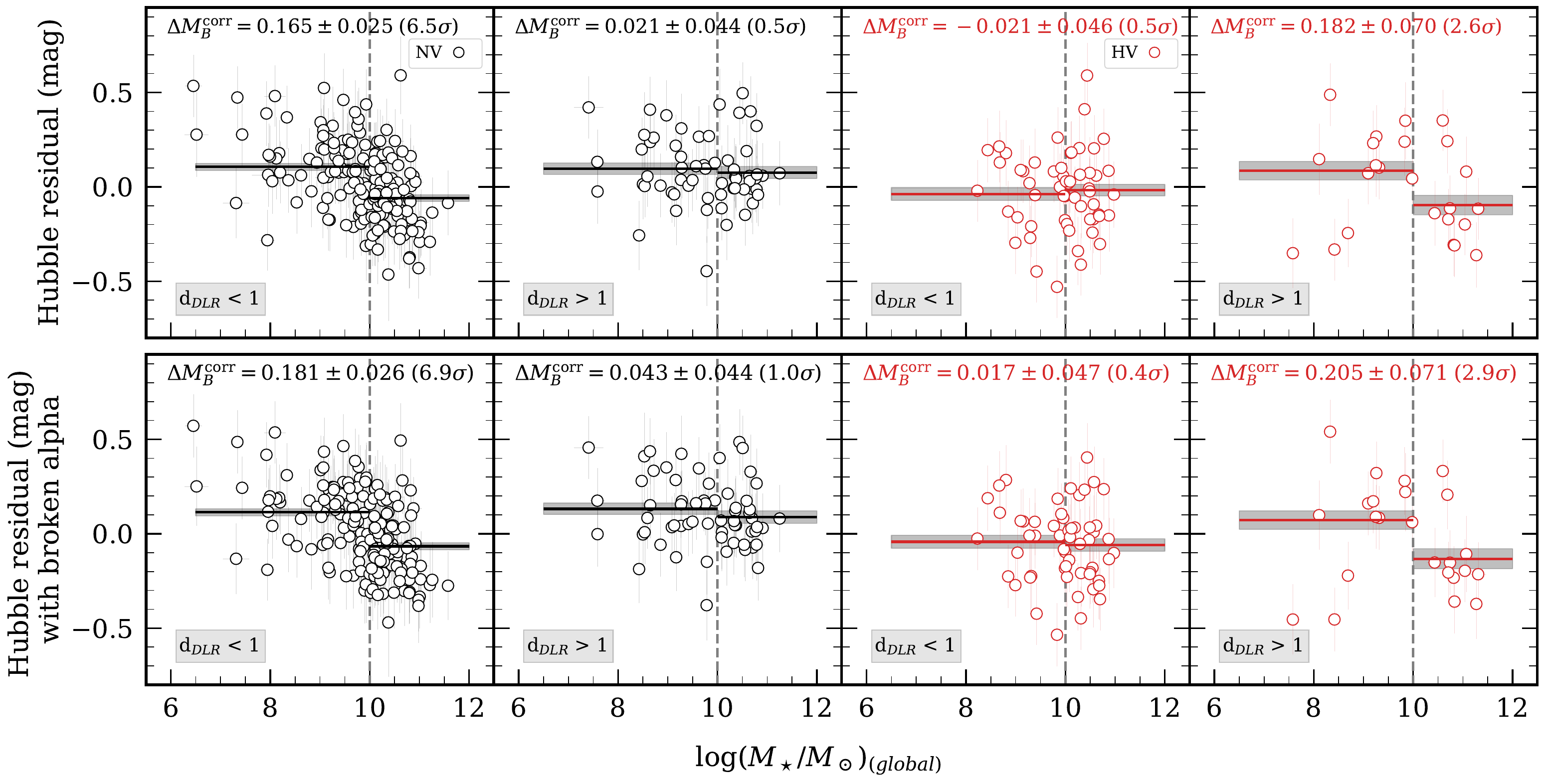}
\caption{Hubble residual plots without and with the broken alpha standardisation as a function of host galaxy mass for HV and NV SNe Ia split by $d_{DLR}$. Black and red dots represent the NV and HV SNe Ia, respectively. The vertical dashed line represents the criterion (log(${M}_\star$/${M}_\odot$) = 10 for the global mass samples). In each plot horizontal dashed lines represent the weighted mean of the Hubble residuals, respectively, where the shaded regions shows the one-sigma uncertainty of the weighted means. The difference of the weighted averages between the low and high mass samples, along with the significances are shown on the label.}
\label{fig:Hr_mass_parametered}
\end{figure*}

\begin{figure*}
\centering
\includegraphics[width=\linewidth]{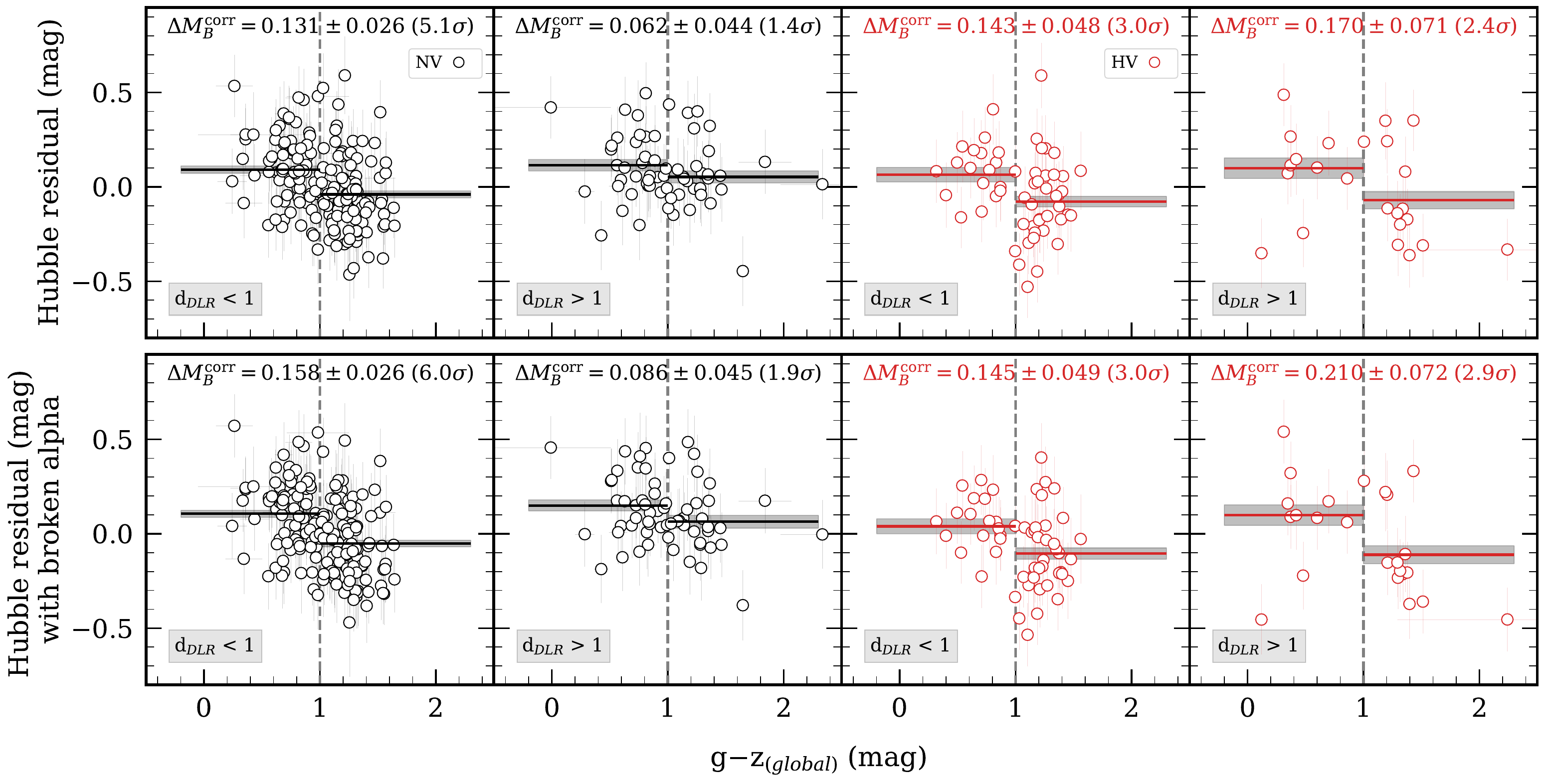}
\caption{Same as Fig.~\ref{fig:Hr_mass_parametered} but as a function of global colour (g$-$z) for HV and NV SNe Ia split by $d_{DLR}$. The vertical dashed line represents the criterion ($g - z$ = 1) used to separate the low and high global colour samples.}
\label{fig:Hr_color_parametered}
\end{figure*}

Considering the differences in sample sizes, we test the robustness of the observed mass step by randomly selecting 70 NV SNe Ia (from the original 207), matching the number of HV SNe Ia. The random draws are weighted to reproduce the same low- and high-mass host proportions as in the HV sample. After $10^5$ resamplings, we find that without the broken-alpha standardisation, only 0.1 per cent of realizations yield a mass step with a significance equal to or smaller than that observed in the HV sample (1.1$\sigma$), with a mean significance of 3.6$\sigma$. The same analysis with the broken-alpha standardisation gives a consistent result: 0.3 per cent of realizations fall below the HV value (1.9$\sigma$), with a mean significance of 3.8$\sigma$. These tests indicate that the weaker HV mass step cannot be explained by the smaller sample size, supporting the conclusion that the mass step is genuinely absent or significantly reduced in HV SNe Ia.

We also investigated the dependence of SN Ia HRs on the global rest frame $g-z$ colour of their host galaxies. In Fig.~\ref{fig:HR_mass} we show the HRs versus the host rest-frame $g-z$ colour split at $g-z =$ 1. The NV SNe Ia show a host-colour HR step of 0.142 $\pm$ 0.024 mag (5.9$\sigma$) and 0.168 $\pm$ 0.025 mag (6.8$\sigma$) without and with broken-alpha standardisation, respectively. In contrast to the host galaxy mass, the HV SNe Ia also show a significant step of  0.158 $\pm$ 0.042 mag (3.8$\sigma$) without broken-alpha, and  0.168 $\pm$ 0.043 mag (4$\sigma$) with broken-alpha. The NV and HV colour steps are statistically consistent.

We also repeat the analysis using local rest-frame $g-z$ colour instead of the global colour. For NV SNe Ia, the step size is similar but the significance is slightly higher, with values of 0.147 $\pm$ 0.024 mag (6.2$\sigma$) without broken-alpha and 0.170 $\pm$ 0.025 mag (6.9$\sigma$) with broken-alpha. For HV SNe Ia, both the size and significance of the step decrease compared to the global colour, with values of 0.129 $\pm$ 0.041 mag (3.1$\sigma$) without broken-alpha and 0.137 $\pm$ 0.042 mag (3.3$\sigma$) with broken-alpha.

\subsection{HRs for host mass/colour and $d_{DLR}$}
\label{sec:dDLR}

In Fig.~\ref{fig:Hr_mass_parametered}, we show the HRs (again without and with broken-alpha standarisation) as a function of global mass split into those that have $d_{DLR}$ < 1 and those that have $d_{DLR}$ > 1. For small $d_{DLR}$, the NV sample shows a global mass step with a significance at 6.5$\sigma$ and a step of 0.165 $\pm$ 0.025 mag without broken-alpha standardisation. No significant step is observed for $d_{DLR}$ > 1 for NV SNe Ia and nor for the HV SNe Ia sample at small or large $d_{DLR}$.  \citet{toy2025} showed that the overall size of the stellar mass step (not split into NV and HV SNe Ia) depends on the galactocentric distance, where for $d_{DLR}$ < 1, a significance of 6.9$\sigma$ with a mass step of 0.100 $\pm$ 0.014 mag was observed. No significant step was observed for $d_{DLR}$ > 1. Our result of a mass step for the NV of 0.165 $\pm$ 0.025 mag (6.5$\sigma$) is consistent with this but we additionally find that the size of the step increases when only considering the NV events, suggesting that the mass step for $d_{DLR}$ < 1 may be specifically driven by NV SNe Ia.

We repeat the analysis of the samples split by $d_{DLR}$, with respect to global colour of the host galaxies, instead of mass, in Fig.~\ref{fig:Hr_color_parametered}. Following \citet{Kelsey2021,Kelsey2023}, \citet{toy2025} showed that the host galaxy rest-frame $U - R$ colour split at 1 shows a significant step for only the $d_{DLR}$ < 1 sample. Our findings agree with this, where we find a 5.1$\sigma$ significance with 0.131 $\pm$ 0.026 step for NV SNe Ia for the $d_{DLR}$ < 1 sample in our host galaxy rest-frame $g-z$ analysis. Interestingly, HV SNe Ia also show a weak trend for the $d_{DLR}$ < 1 subsample with 3.0$\sigma$ significance when considering colour, where no significant mass step was seen for this sample.

To further quantify the differences observed in both the low and high $d_{DLR}$ samples, we performed KS tests that compared NV and HV SNe Ia within each sub-sample separately. For the low $d_{DLR}$ sample, there is no strong evidence that the distributions of global and local galaxy mass, global and local rest-frame colour, or SALT2 $x_1$ differ between NV and HV SNe Ia. However, in this low $d_{DLR}$ subsample, we find a difference in the SALT2 $c$ parameter between NV and HV SNe Ia, with a p-value of 0.0156, indicating that HV SNe Ia tend to be redder than NV SNe Ia. Similarly, for the high $d_{DLR}$ sample, there is no significant difference between NV and HV SNe Ia for global and local galaxy mass, global and local colour, or $x_1$, but the SALT2 $c$ parameter also shows a significant difference (p-value = 0.0176), again suggesting redder colours for HV SNe Ia. These results are consistent with trends previously reported for HV and NV SNe Ia in terms of the SALT2 $c$ parameter \citep{wang2013,burgaz2025a}.

\section{Discussion}
\label{sec:discussion}

In this work, we have investigated the relation between SN Ia HRs, obtained from light-curve fits to the ZTF SN Ia sample, and the intrinsic \SiII\ 6355 velocities measured from SN Ia spectra around maximum light.  We also investigated the connection between the extrinsic host galaxy `mass' or `colour' HR step and the intrinsic \SiII\ velocities of the SNe Ia. 

\subsection{A mass step driven by the NV sample}

A key result is that we have identified strong mass steps of 0.141 -- 0.177 mag at 6 -- 7$\sigma$ significance in the NV sample, depending on whether local or global masses, or standardisation without or with broken alpha, are included. For the HV sample, the corresponding steps are smaller (0.046 -- 0.089 mag) and of low significance. The difference between the global NV and HV mass steps is at a modest significance of $\sim$2.2$\sigma$ without broken-alpha and $\sim$1.7$\sigma$ with broken-alpha. While the global NV -- HV difference is only marginally significant, the stronger NV trend compared to HV could indicate that the origin of the mass step may be at least partly connected to intrinsic SN Ia properties, such as the explosion velocity as measured here.

Similar colour steps exist of 0.142 -- 0.170 mag at 6 -- 7$\sigma$ significance for the NV sample. The HV sample also shows colour steps of 0.129 -- 0.168 mag, with lower significance (2.5–3.8$\sigma$) possibly due to smaller sample size, but the NV and HV colour-steps are statistically consistent. This could suggest that the mass step is driven by something more intrinsic to the SNe Ia than the colour step, since it is seen only for NV SNe, and that the colour of the SN environment at the time of explosion is less directly related to the origin of the diversity of SN Ia velocities. 

\begin{figure}
\centering
\includegraphics[width=\columnwidth]{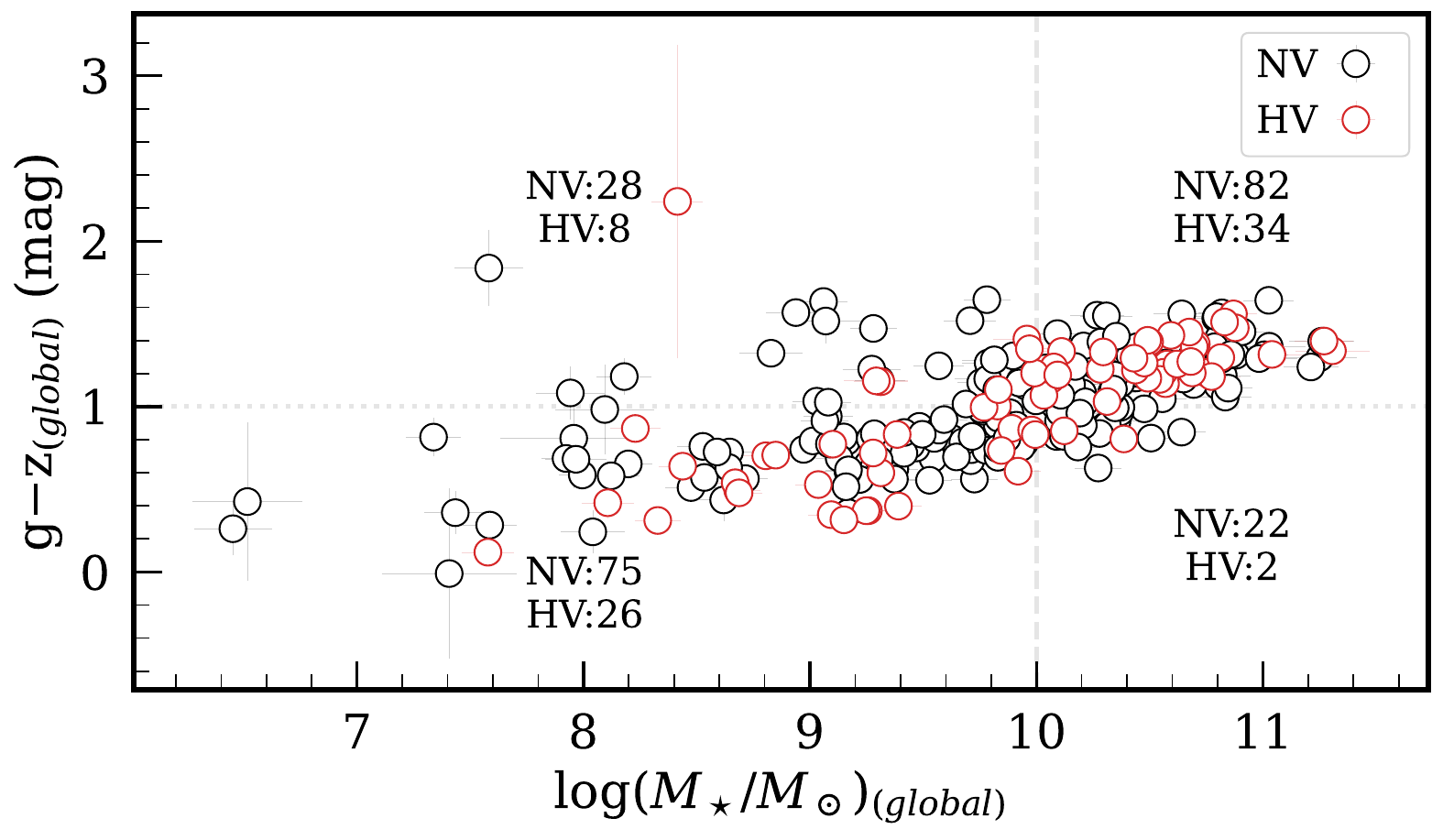}
\caption{Distribution of NV and HV SNe Ia across host galaxy mass and global color. HV and NV SNe Ia are represented with red and black circles, respectively. The vertical dashed line at log(${M}_\star$/${M}_\odot$) = 10 marks the division between low- and high-mass galaxies, and the horizontal dotted line at $g-z$ = 1 marks the division between blue and red hosts.}
\label{fig:gz_mass}
\end{figure}

The reason that we see a colour and not a mass step for the HV is subtle and is driven by small numbers of SNe Ia moving from high mass to blue colour or red colour to low mass, which causes them to move `sides' of the step. For reference, the NV counts are 28 (low-mass red), 75 (low-mass blue), 82 (high-mass red), and 22 (high-mass blue), while the corresponding HV counts are 8, 26, 34, and 2, respectively (Fig.~\ref{fig:gz_mass}). In three quadrants (low-mass red, low-mass blue, and high-mass red), the HV-to-NV ratio is consistently $\sim$1:3, but in the high-mass blue bin, we find only two HV events. Under the assumption of a uniform HV fraction, we would expect 7.4 $\pm$ 2.7 HV SNe Ia in this bin, corresponding to a $\sim$2$\sigma$ difference from the observed number of two SNe. This suggests that HV SNe Ia may be less common in high-mass blue galaxies. However, the statistical significance is only modest, and larger samples are required to confirm this effect. Therefore, it is difficult to determine the reason for the differences seen in the amplitudes and significances of the mass and colour steps.

Given the size of the volume-limited ZTF DR2 sample, as well as the inclusion of SNe Ia in host galaxies without spectroscopic redshifts, it contains more SNe Ia in low-mass host galaxies than previous samples. \citet{burgaz2025b} ran a series of analyses on HV SNe Ia in low-mass galaxies and found that the occurrence rate is similar to that of NV SNe Ia. This was an update to previous results that suggested a higher occurrence in high-mass hosts that was likely due to sample biases. In this analysis, there is only one HV SNe Ia with a host galaxy mass, log(${M}_\star$/${M}_\odot$), smaller than 8, while there are 12 NV SNe Ia. To test how these SNe Ia in the lowest mass hosts might impact our results, we removed the SNe Ia with log(${M}_\star$/${M}_\odot$) < 8, which dropped the mass step to 0.136 $\pm$ 0.025 mag (5.5$\sigma$) for NV SNe Ia, but still remains significant. Since the significance change is around 0.7$\sigma$, a similar increase in HV SNe Ia, if they had the same number of objects in the log(${M}_\star$/${M}_\odot$) < 8 area would still result in a significance lower than 2$\sigma$.

For the mass step when the ZTF SN Ia sample is split into HV and NV samples, it appears that the difference is driven by a shift of the mean HR in low mass galaxies towards a more positive value for NV SNe, with both the NV and HV SNe in high mass galaxies having similar HRs, regardless of the use of the broken alpha standardisation or not, or global or local stellar mass. For the HRs with respect to galaxy colour, the mean HR for NV SNe in blue and red galaxies is similar to that of NV SNe in low and high mass galaxies, respectively. For HV SNe, the appearance of a colour step, where no global mass step is seen, is driven by a shift towards more positive HRs in blue galaxies and towards more negative HRs in red galaxies. These HV shifts correspond to modest ($\sim$2.5–3.8$\sigma$) detections, which are not drastically different in significance from NV results in certain bins. As discussed in \cite{burgaz2025a}, there is a small difference in the underlying SALT2 \textit{$x_1$} and \textit{c} distributions of the NV and HV samples, with the HV sample having marginally narrower light curves and redder colours, on average. \cite{wang2013} showed for a sample of SNe Ia from the galaxy-targeted Lick Observatory Supernova Search that HV SNe Ia preferentially resided in the inner regions of their host galaxies, as measured by the semi-major axis at the 25 mag arcsec$^2$ isophote in the \textit{B} band. However, this is not seen in this untargeted ZTF SN Ia sample, where there is a constant ratio of $\sim$30 \% HV to NV up to a $d_{DLR}$ of $\sim$2 \citep{burgaz2025b}. The $d_{DLR}$ corrects for the impact of inclination that is not accounted for when using the semi-major axis, which along with the targeted vs.~untargeted nature of the survey, likely explains the differences in the results. 

\subsection{Connection to host galaxy galactocentric distance}

We have also performed a comparison of SNe Ia with different peak velocities occurring in the central regions of their hosts ($d_{DLR}$ < 1) and those in the outskirts ($d_{DLR}$ > 1). We confirmed the relation between mass step and $d_{DLR}$ of \cite{toy2025} but further showed that this relation is driven by the NV SNe Ia, with a strong and statistically significant mass step of 0.165 $\pm$ 0.025 mag (6.5$\sigma$) for the NV sample at small $d_{DLR}$. In contrast, when NV SNe occur in galaxy outskirts ($d_{DLR}$ > 1), the mass step disappears entirely. 
A similar trend emerges when examining global and local host restframe colour (\textit{g-z}), where NV SNe show a 5$\sigma$ colour step at $d_{DLR}$ < 1, but no significant step in the outskirts. 

For HV SNe Ia, the central-region mass step is consistent with zero ($-0.021 \pm 0.027$ mag). This yields a NV -- HV contrast of 3.1 -- 3.6$\sigma$, representing the most significant subtype difference we find. However, HV SNe exhibit non-zero steps of $\sim$ 0.18 -- 0.21 mag in the outskirts, with 2.6 -- 2.9$\sigma$ significance. This suggests for HV SNe Ia that while the residuals of their luminosities are unaffected by stellar mass, they may still retain some sensitivity to local stellar population age, metallicity, or dust content, as traced by host colour \citep{ginolin2025b}. The fact that a weak colour step exists in both the inner and outer regions of the host galaxies of HV SNe (although less pronounced than for NV SN samples at small $d_{DLR}$) suggests that HRs of HV SNe are subtly linked to their galaxy colours but this is relatively independent of $d_{DLR}$.

The properties of galaxies as a function of radial distance have been investigated, with the identified trends depending on galaxy type. For both early type/elliptical and late type/spiral galaxies, the stellar metallicity is found to decrease slightly with increasing radial distance \citep{gonzalezdelgado2015,goddard2017}. However, the dependence of the ages of stars with radial distances is different in different galaxy types, stars in the outer regions of early-type galaxies are likely the same or marginally older than the inner regions \citep{goddard2017}. In late-type galaxies, the inner regions dominated by bulges are likely older than the outer regions. Dust gradients also exist within galaxies, with all galaxy types showing increased extinction due to dust towards their centres \citep{gonzalezdelgado2015,goddard2017} and several studies have proposed that dust may contribute significantly to the mass step \citep{Brout2021,popovic2023}. However, \cite{toy2025} investigated if the difference in the mass step they observed between inner and outer regions of their host galaxies and found varying total-to-selective extinction, R$_V$ was insufficient to explain it. Therefore, it remains an open question what is driving these relations. However, the fact that the galaxy steps sizes differ when the sample is split based on intrinsic velocity of the SN ejecta and the distance from the centre of the host galaxy, suggests that the origin must be at least somewhat intrinsic to the SN and dependent on progenitor scenario or the specifics of the explosion mechanism. 

\subsection{The connection between galaxy HR steps and intrinsic SN properties}

Our findings imply that the widely reported HR steps with host galaxy mass and colour are not universal across all SNe Ia, but instead are strongly dependent on at least one intrinsic property of SNe Ia, their explosion velocities as measured at peak. Although the overall difference in the mass step between NV and HV SNe Ia is only marginal ($\sim$2.2$\sigma$), the evidence for subtype dependence is stronger when focusing on SNe located in the central regions of their hosts ($d_{DLR}<1$), where the contrast reaches 3.1--3.6$\sigma$. Moreover, the way in which the mass and colour steps vary with galactocentric distance differs between the two velocity subtypes. Hence, the subtype and location dependence raises the possibility that dust is not the sole driver of the mass and colour steps. Additional factors such as intrinsic SN Ia properties or local environments may play a role, and correlations between progenitor age, dust properties, and explosion velocity could also contribute.

The HV SNe appear to more uniform across a range of stellar masses and colours, but still show modest ($\sim$2.5 -- 3$\sigma$) steps in some bins, while the luminosities of the NV SNe show a dependence on both mass and colour. Previous studies such as \cite{Rigault2013} and \cite{Childress2013} have suggested that the mass/colour step may be caused by differences in the luminosity of SNe Ia from young and old populations or two different progenitor channels that have different mean ages and luminosities. However, \cite{wiseman2023} suggested that it is likely that a luminosity step with progenitor age, as well as a total-to-selective dust extinction ratio, R$_V$, that changes with galaxy age, that drives the differences in HR residuals in different galaxy environments. Our results are in agreement with \cite{wiseman2023} and show hints of intrinsic differences between HV and NV SNe Ia. If that is the case, the contrast could be linked to progenitor age together with dust effects, possibly including different dust laws for the two subtypes.

However, a completely intrinsic origin for the galaxy steps cannot be ruled out because if, for example, there were subtle metallicity differences in the progenitor stars (due to different birth environments), then this could cause differences in the observed properties of the SNe Ia and hence the HRs. This has been suggested theoretically even within specific explosion scenarios, such as the double-detonations of sub-Chandrasekhar mass white dwarfs where differences in nucleosynthetic yields are seen when just the metallicity is varied \citep[e.g.,][]{shen2018,leung2020,gronow2021}. While these studies do not directly predict Hubble residuals, such luminosity shifts would appear as residual offsets after standardisation if present in real populations, as also suggested by observational trends between metallicity and HRs \citep[e.g.,][]{Childress2013, pan2014}.

The way in which NV and HV SNe Ia respond differently to their environments may alternatively hint that they come from different progenitor systems or explosion mechanisms. The strong correlation between HRs and local host properties for NV SNe, particularly in central, metal-rich, and older stellar populations, points to a scenario where NV SNe preferentially originate from older, higher metallicity progenitor systems. This is consistent with a longer time until explosion for the progenitor scenario (e.g., double-degenerate systems or single-degenerate systems with longer delay times), where the progenitor white dwarf(s) evolve in environments shaped by metallicity and age \citep[e.g.,][]{zhengwei2023}. However, the relatively weak or completely absent environmental dependence for HV SNe, especially the lack of a mass step at any galactocentric radius, implies that their progenitor systems may be more uniform and/or possibly younger, arising from a different or less environmentally sensitive channel. This could reflect a preference for either a star-forming or an intermediate-age environment, or lower metallicity.

For NV SNe, global host properties only predict SN luminosity residuals when the SN arises from the central galaxy region, supporting a local-environment origin of the mass and colour step. For HV SNe, environmental trends are weaker but not always absent, and some subsets show non-zero steps at modest significance, which could reflect intrinsic differences between NV and HV SNe Ia, such as variations in progenitor channels or explosion physics.

These results highlight the importance of incorporating local environmental context and spectroscopic information into SNe Ia standardisation for precision cosmology.

\section{Conclusions}
\label{sec:conclusions}

Using the spectroscopically classified SNe Ia from the volume-limited sample of ZTF DR2 at peak light, we have analyzed the NV and HV SNe Ia through their HRs and studied the influence of light-curve parameters and host galaxy separation on the estimated mass and colour steps. The main conclusions are:

\begin{enumerate}
      \item HV and NV SNe Ia show no significant differences in HRs in most cases. However, under the specific case of a broken alpha standardisation and when all redshift sources are included, a 3.7$\sigma$ significance is found. Excluding template-matched redshifts reduces significance, suggesting that the selection of redshift sources and sample size influence the results.
      \item Significant trends in HR are found within the NV SNe Ia sample when split by $x_1$, $c$, host mass, $d_{DLR}$ and local/global environment properties when using broken-alpha standardisation, with the strongest trends reaching up to 6.9$\sigma$ when considering environmental colour. HV SNe Ia show weaker internal trends overall but do exhibit modest signals (typically $\sim$2.5 -- 3$\sigma$) in some subsets, including local/global colour and outer regions. 
      \item We found a highly significant (6.3$\sigma$) mass step in NV SNe Ia, while HV SNe Ia show no significant mass step. Quantitatively, the global NV and HV mass steps differ at a modest $\sim$2.2$\sigma$ (no broken-alpha) or $\sim$1.7$\sigma$ (broken-alpha) difference.
      \item We observe a $0.142 \pm 0.024$ mag ($5.9\sigma$) colour step in NV SNe Ia, and a $0.158 \pm 0.042$ mag ($3.8\sigma$) colour step in HV SNe Ia. Even though the HV SNe Ia colour step is larger, its significance appear modest possibly due to the smaller sample size. Overall, NV and HV colour steps are statistically consistent.
      \item The mass-step is present only for NV SNe Ia located near the centres of their host galaxies ($d_{DLR}$). For NV SNe in the outskirts, the mass step disappears entirely. On the other hand, HV SNe Ia show no central-region mass step but display larger steps ($\sim$0.18 -- 0.21 mag) at $\sim$2.6 -- 2.9$\sigma$ in outer regions. The most significant subtype difference occurs for central events ($d_{DLR}<1$), where NV show a mass step while HV are consistent with none, corresponding to a 3.1 -- 3.6$\sigma$ difference between two subtypes.
      \item A KS test shows that HV and NV SNe Ia located near galaxy centres ($d_{DLR}$ < 1) differ in SALT2 colour ($c$), with HV SNe showing redder values, which supports the idea that HV SNe are either dustier or have intrinsically different colours, consistent with differences in progenitor systems or explosion mechanisms.
      \item A similarly strong dependence on host rest-frame global colour ($g-z$) is observed for NV SNe only when the SNe occur in the central regions of the galaxy are considered, suggesting this could be driven by local stellar population properties, such as metallicity, age or dust rather than global host mass or colour. We therefore emphasize central environments as the likely drivers of the NV trend.  
      \item HV SNe Ia show a modest host galaxy colour step in both central and outer regions (3.0$\sigma$ and 2.4$\sigma$, respectively), suggesting a possible sensitivity to progenitor age or dust, but not sufficiently strong to be used as a reliable correction. These results are statistically consistent with the NV colour steps.
   \end{enumerate}

How NV and HV SNe Ia respond differently to host galaxy environment suggests an important connection between the intrinsic properties of the SN explosion (e.g., explosion energy) and the environments within which the SNe Ia were born or explode. However, because the global NV-HV mass-step difference is at modest $\sim$2.2$\sigma$, our interpretation relies primarily on the stronger central region result (3.1--3.6$\sigma$). We find a strong correlation between Hubble residuals and host properties for NV SNe, particularly in central, likely metal-rich and older stellar populations. This suggests that while NV SNe Ia are environmentally sensitive, HV SNe Ia are not, possibly because they represent a more homogeneous population.

Our results imply that SNe Ia standardisation needs to account for both spectroscopic subtype and local environment. Uniform application of mass or colour steps may introduce systematic errors if these variables are not carefully considered. We recommend reporting the measured steps with uncertainties for each subtype and galactocentric bin, and adopting subtype- and location-aware corrections (e.g., $d_{DLR}$-weighted) rather than a single universal step. We find that the correlation between global host mass and SN luminosity residuals is strongest when the SN is located near the galaxy centre, suggesting that host galaxy properties may not be equally informative at all galactocentric distances. This reinforces the value of spatially resolved environmental metrics in future cosmological surveys. In future SN Ia cosmology analyses, incorporating both local photometry and/or spectroscopy and SN subtype may significantly improve the scatter in precision cosmology. In particular, spatially resolved spectroscopy to constrain environmental properties such as metallicity and stellar population age could help clarify the role of host environment, while galactocentric distance may serve as a useful weighting factor in SN standardisation to reduce residual scatter.

\begin{acknowledgements}
Based on observations obtained with the Samuel Oschin Telescope 48-inch and the 60-inch Telescope at the Palomar Observatory as part of the Zwicky Transient Facility project. ZTF is supported by the National Science Foundation under Grants No. AST-1440341 and AST-2034437 and a collaboration including current partners Caltech, IPAC, the Weizmann Institute of Science, the Oskar Klein Center at Stockholm University, the University of Maryland, Deutsches Elektronen-Synchrotron and Humboldt University, the TANGO Consortium of Taiwan, the University of Wisconsin at Milwaukee, Trinity College Dublin, Lawrence Livermore National Laboratories, IN2P3, University of Warwick, Ruhr University Bochum, Northwestern University and former partners the University of Washington, Los Alamos National Laboratories, and Lawrence Berkeley National Laboratories. Operations are conducted by COO, IPAC, and UW. SED Machine is based upon work supported by the National Science Foundation under Grant No. 1106171.

U.B, K.M., and T.E.M.B. are funded by Horizon Europe ERC grant no. 101125877. JHT is supported by the H2020 European Research Council grant no. 758638.

L.G. acknowledges financial support from AGAUR, CSIC, MCIN and AEI 10.13039/501100011033 under projects PID2023-151307NB-I00, PIE 20215AT016, CEX2020-001058-M, ILINK23001, COOPB2304, and 2021-SGR-01270.

This project has received funding from the European Research Council (ERC) under the European Union's Horizon 2020 research and innovation program (grant agreement n 759194 - USNAC).

Y.-L.K. was supported by the Lee Wonchul Fellowship, funded through the BK21 Fostering Outstanding Universities for Research (FOUR) Program (grant No. 4120200513819) and the National Research Foundation of Korea to the Center for Galaxy Evolution Research (RS-2022-NR070872, RS-2022-NR070525).

This research used resources of the National Energy Research Scientific Computing Center (NERSC), a Department of Energy User Facility using NERSC award HEP-ERCAP0033784. 

This work has been supported by the research project grant “Understanding the Dynamic Universe” funded by the Knut and Alice Wallenberg Foundation under Dnr KAW 2018.0067. AG acknowledges support from  {\em Vetenskapsr\aa det}, the Swedish Research Council, project 2020-03444, as well as the Swedish National Space Agency, under Dnr 2023-00226.

NR is supported by DoE award \#DE-SC0025599. Zwicky Transient Facility access for NR was supported by Northwestern University and the Center for Interdisciplinary Exploration and Research in Astrophysics (CIERA)

\end{acknowledgements}

\bibliographystyle{aa}
\bibliography{aanda} 

\begin{appendix}

\section{Sample selection effect}
\label{appASample}

\begin{figure*}
\centering
\includegraphics[width=16.8cm]{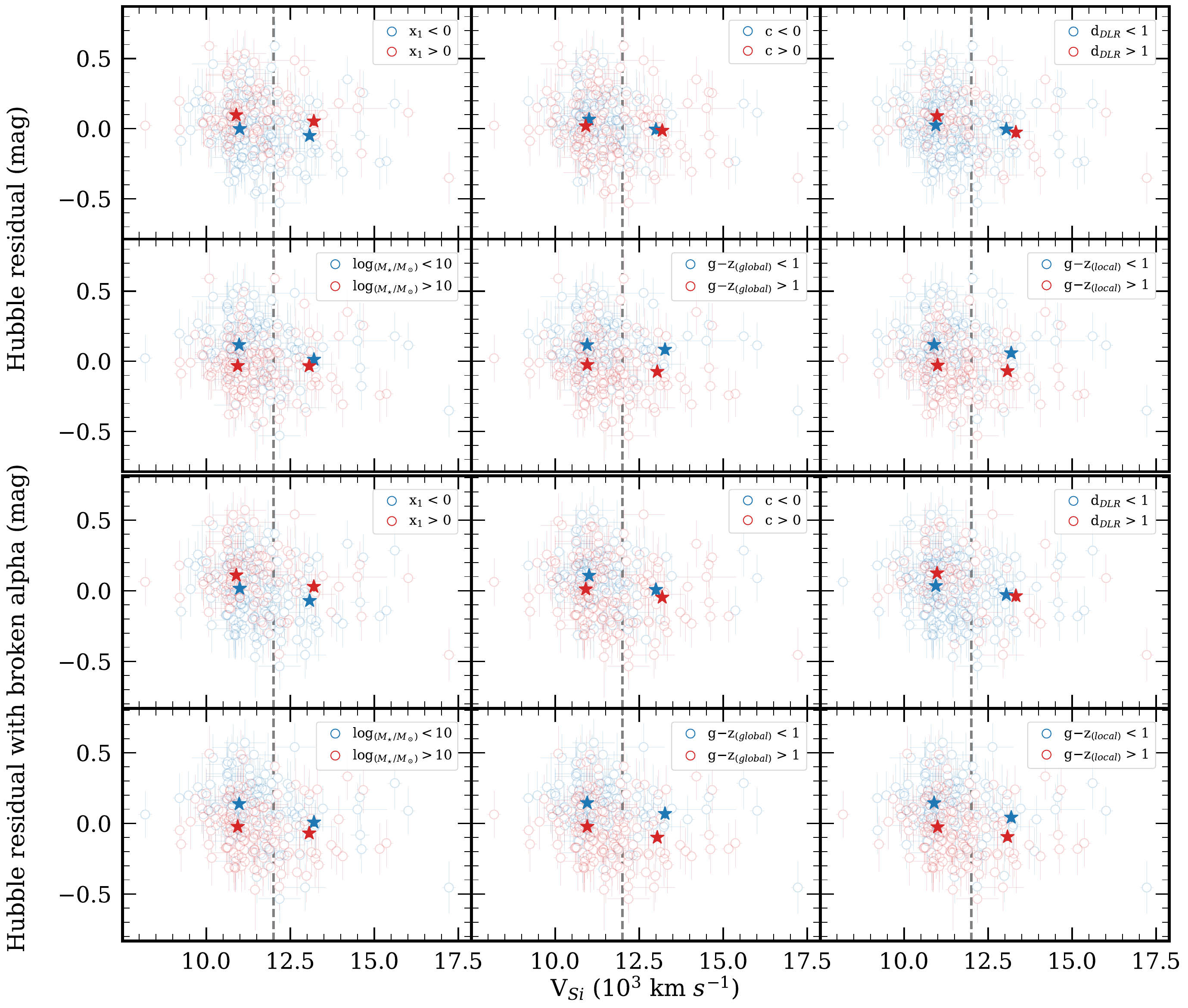}
\caption{Same as Figure  Fig.~\ref{fig:HR_HV_NV}, but showing the full distribution of Hubble residuals without (top two rows) and with (bottom two rows) the broken-alpha standardisation of HV and NV SNe Ia.}
\label{fig:appA}
\end{figure*}

To assess the potential effects of a selection bias, we do a series of analysis in this work. We first explore the sample selection effect coming from the chosen velocity separation value for the NV and HV SNe Ia. While in this work we mainly adapted 12,000 \kms, in literature there are different selections such as 11,000 \kms\ in \citet{Siebert2020} and 11,800 \kms\ in \citet{Dettman2021}. Hence, we also adapted 11,800 \kms\ as threshold and did the same analysis.

Secondly, in past studies, 99a-like and all 04gs-like SN Ia were probably included in normal SN Ia samples. Hence, in order to check the effect of removing these sub-types, we made the analysis by including 13 04gs-like SN Ia, where 9 would be NV and 4 would be HV and 34 99aa-like SN Ia, where 28 would be NV and 6 would HV, following same classification system as normals. 

Lastly, we check the effect of selected phase threshold. In this work the peak sample is defined as within 5 days since maximum light, however, we also provide a more strict definition of the peak sample as within 3 days.

\begin{figure*}
\centering
\includegraphics[width=14cm]{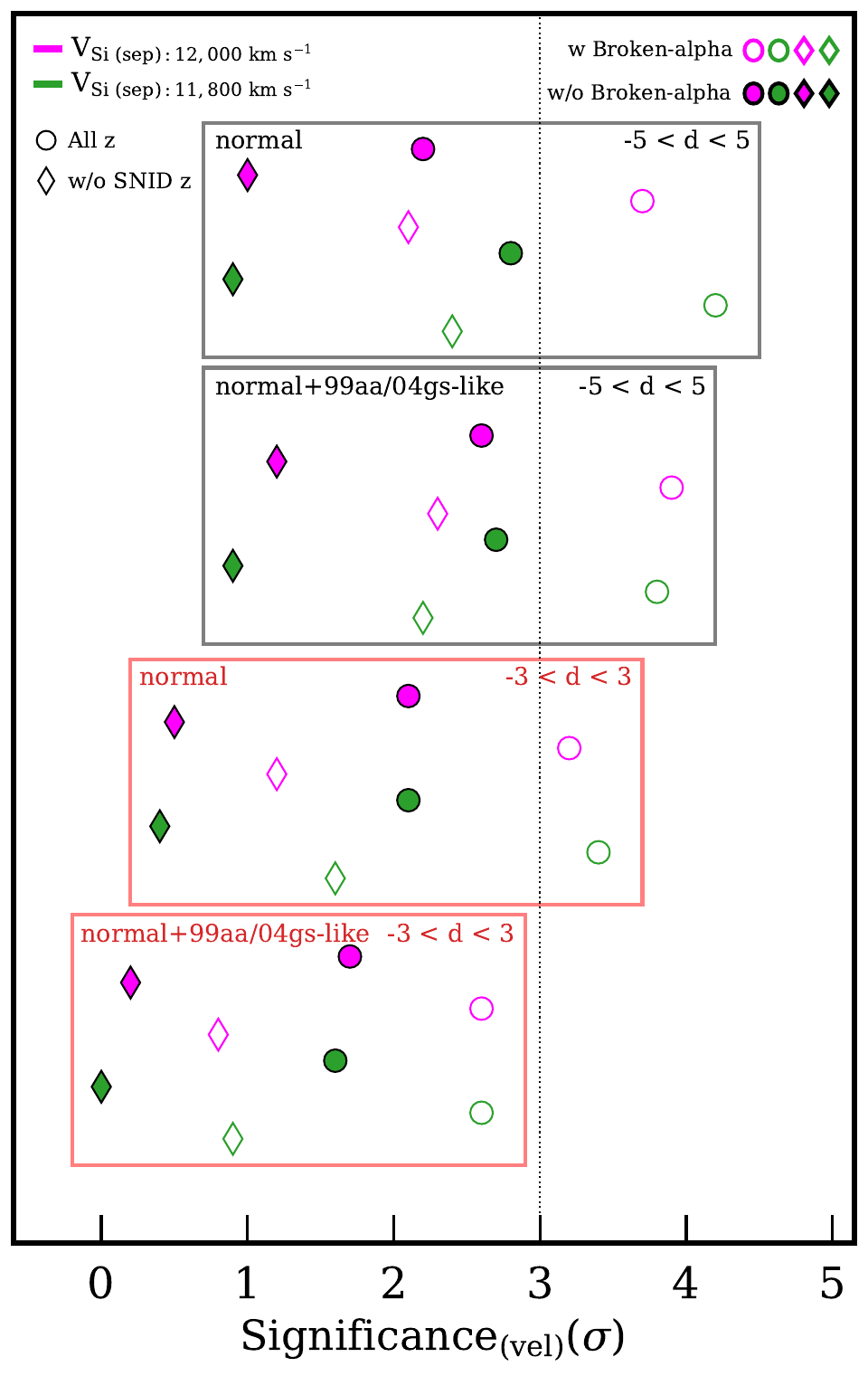}
\caption{Significances of the weighted average Hubble residual differences as a function of velocity, following the same general analysis as Fig.~\ref{fig:HR_vel_12000}. The circles represent samples that include all redshift sources, while the diamonds represent samples excluding template-matched redshift sources. Filled symbols indicate Hubble residuals standardized without the broken-alpha correction, and unfilled symbols show results with the broken-alpha correction. Magenta and green colours correspond to velocity splits at 12,000 \kms\ and 11,800 \kms, respectively. The dashed vertical line marks the significance threshold. The black boxes show the results for samples in which the peak phase is within 5 days and the red boxes show the results in which the peak phase is within 3 days. The sub-types included in each analysis are noted within the boxes.}
\label{fig:appA1}
\end{figure*}

In Fig.~\ref{fig:appA1}, we show the significances of the weighted average Hubble residual differences as a function of velocity (similar to Fig.~\ref{fig:HR_vel_12000}), including all the selection effects considered. Regardless of the case there seems to be a none to weak relation, where the significance between the NV and HV SNe Ia seems to be mostly below 3$\sigma$ level. Changing the velocity separation threshold from 12,000 \kms\ to 11,800 \kms\ slightly increasing the significances that include only normal SNe Ia. This effect seems to diminish once the 99aa-like and 04gs-like SNe are included in the samples. In general, inclusion of 99aa-like and 04gs-like SNe does not seem to be effecting anything significantly. We also clearly see that in all cases inclusion of template matched SNe to that sample increases the significances. Exclusion of those SNe would introduce a bias towards the high-mass galaxies since in most cases the SN without a redshift source is low-luminosity, hence a low-mass galaxy. That is why in all the analysis done in this work before and from this point includes all redshift sources. Lastly, using the broken-alpha standardisation from \citet{ginolin2025b} increases the observed significances.

\begin{figure*}
\centering
\includegraphics[width=14cm]{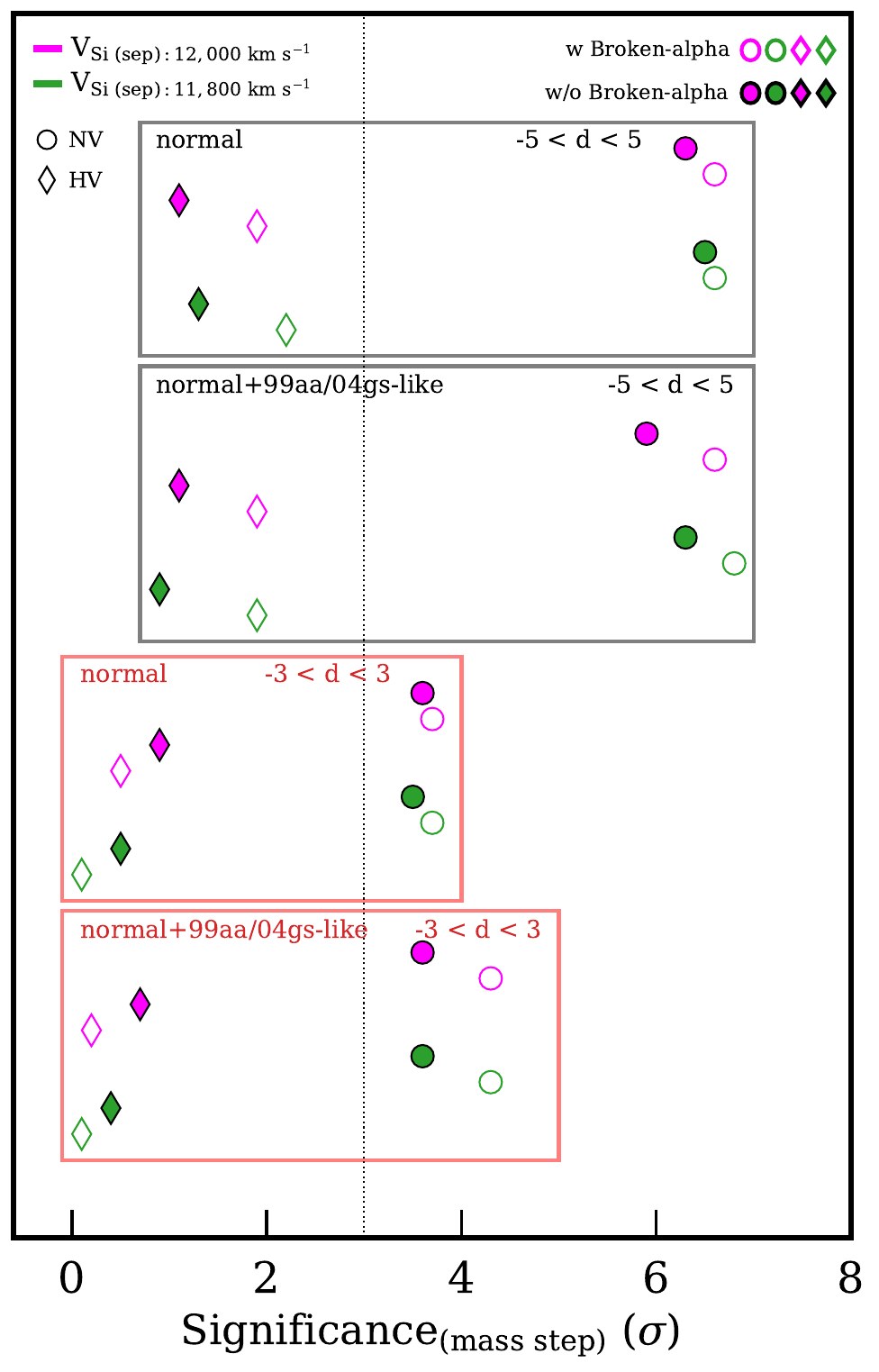}
\caption{Significances of the weighted average Hubble residual differences as a function of mass following the same general analysis as Fig.~\ref{fig:HR_mass}. The circles represent the NV SNe Ia and the diamonds represent the HV SNe Ia. Filled symbols indicate Hubble residuals standardized without the broken-alpha correction, and unfilled symbols show results with the broken-alpha correction. Magenta and green colours correspond to velocity splits at 12,000 \kms\ and 11,800 \kms, respectively. The dashed vertical line marks the significance threshold. The black boxes show the results for samples in which the peak phase is within 5 days and the red boxes show the results in which the peak phase is within 3 days. The sub-types included in each analysis are noted within the boxes.}
\label{fig:appA2}
\end{figure*}

\begin{figure*}
\centering
\includegraphics[width=14cm]{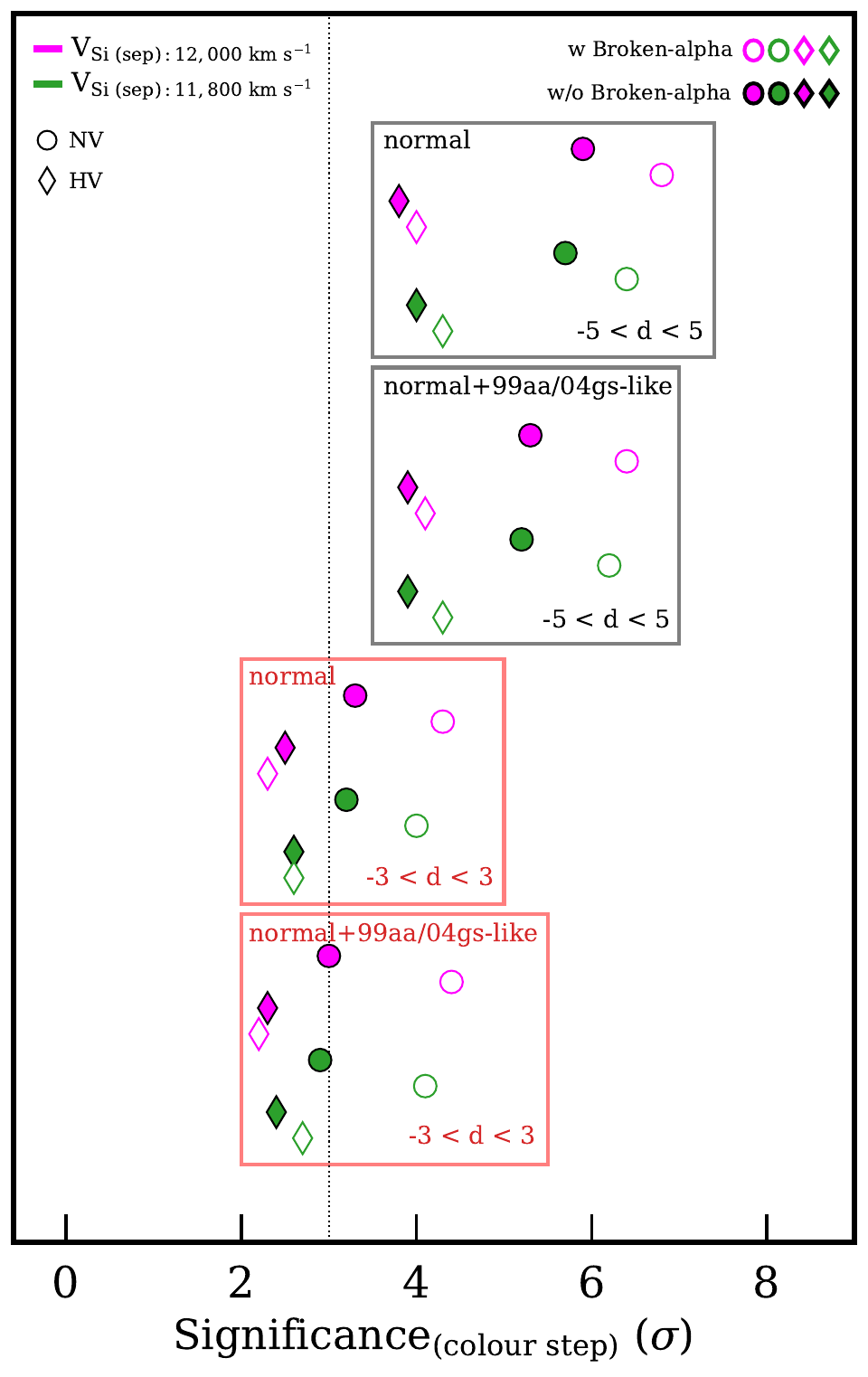}
\caption{Significances of the weighted average Hubble residual differences as a function of global colour ($g-z$), following the same general analysis as Fig.~\ref{fig:HR_mass}. The circles represent the NV SNe Ia and the diamonds represent the HV SNe Ia. Filled symbols indicate Hubble residuals standardized without the broken-alpha correction, and unfilled symbols show results with the broken-alpha correction. Magenta and green colours correspond to velocity splits at 12,000 \kms\ and 11,800 \kms, respectively. The dashed vertical line marks the significance threshold. The black boxes show the results for samples in which the peak phase is within 5 days and the red boxes show the results in which the peak phase is within 3 days. The sub-types included in each analysis are noted within the boxes.}
\label{fig:appA3}
\end{figure*}

In Fig.~\ref{fig:appA2} and Fig.~\ref{fig:appA3}, we explore the sample selection effect on the mass and colour steps, respectively (similar to Fig.~\ref{fig:HR_mass}). We notice hanging the velocity separation threshold from 12,000 \kms\ to 11,800 \kms\ seems to irrelevant in almost all cases, while using the broken-alpha standardisation seems to increase the significance slightly. Using 11,800 \kms\ also gives out a slightly higher significance difference between the mass-steps of HV and NV SNe Ia (possibly due to increased sample size of HV SNe Ia), at 2.6$\sigma$, where 12,000 \kms\ separation were giving 2.2$\sigma$. Lastly, using a smaller phase yields lower significances in all analysis, however we observe that inclusion of 99aa-like and 04gs-like SNe have no effect to the observed significances. 

\end{appendix}

\end{document}